\documentclass[sigconf]{acmart}
\AtBeginDocument{%
  }

\usepackage{subfigure}
\usepackage{enumitem}

\newcommand{\bbb}{\noindent\textbf}

\newcommand{\alg}{LLM-Sketch}

\begin{document}

\title{\alg{}: Enhancing Network Sketches with LLM}


\author{Yuanpeng Li}
\affiliation{%
  \institution{Peking University \\ Zhongguancun Laboratory}
  \city{Beijing}
  \country{China}
}

\author{Zhen Xu}
\affiliation{%
  \institution{Zhejiang University}
  \city{Hangzhou}
  \country{China}
}

\author{Zongwei Lv}
\affiliation{%
  \institution{Peking University}
  \city{Beijing}
  \country{China}
}

\author{Yannan Hu}
\authornote{Tong Yang (yangtong@pku.edu.cn) and Yannan Hu (huyn@zgclab.edu.cn) are the corresponding authors.}
\affiliation{%
  \institution{Zhongguancun Laboratory}
  \city{Beijing}
  \country{China}
}

\author{Yong Cui}
\affiliation{%
  \institution{Tsinghua University \\ Zhongguancun Laboratory}
  \city{Beijing}
  \country{China}
}

\author{Tong Yang}
\authornotemark[1]
\affiliation{%
  \institution{Peking University \\ Zhongguancun Laboratory}
  \city{Beijing}
  \country{China}
}

\renewcommand{\shortauthors}{Li et al.}

\begin{abstract}
  Network stream mining is fundamental to many network operations. Sketches, as compact data structures that offer low memory overhead with bounded accuracy, have emerged as a promising solution for network stream mining. Recent studies attempt to optimize sketches using machine learning; however, these approaches face the challenges of lacking adaptivity to dynamic networks and incurring high training costs.
In this paper, we propose LLM-Sketch, based on the insight that fields beyond the flow IDs in packet headers can also help infer flow sizes. By using a two-tier data structure and separately recording large and small flows, LLM-Sketch improves accuracy while minimizing memory usage. Furthermore, it leverages fine-tuned large language models (LLMs) to reliably estimate flow sizes. We evaluate LLM-Sketch on three representative tasks, and the results demonstrate that LLM-Sketch outperforms state-of-the-art methods by achieving a \(7.5 \times\) accuracy improvement.

\end{abstract}



\keywords{Network stream mining, Sketches, Flow classification, LLM}

\maketitle

\section{Introduction}

Network stream mining is a cornerstone of modern computer networking, underpinning critical tasks such as DDoS victim detection \cite{poseidon, jaqen}, load balancing \cite{conga, silkroad}, congestion control \cite{hpcc}, and traffic engineering \cite{hedera, aasclepius}. As network scale and traffic volumes continue to grow, ensuring efficient and accurate stream mining at scale becomes increasingly challenging. In response, sketches \cite{bloomfilter, cmsketch, cusketch, csketch} have gained traction for their compact nature and ability to provide small-yet-bounded error under stringent memory constraints, making them well-suited for large-scale network stream mining scenarios.

Despite their appeal, existing sketch solutions often struggle to maintain acceptable error rates in the face of massive-scale networks and highly skewed traffic distributions \cite{cusketch, imcdc}. In practice, a small fraction of large flows typically accounts for the majority of total traffic volume, while many small flows remain numerous yet contribute only modestly. A representative example is the Count-Min Sketch (CMS) \cite{cmsketch}, which updates and queries counters based on hashed flow IDs. Although CMS is simple and memory-efficient, it faces a fundamental trade-off: counters sized for small flows undercount the large ones, while counters sized for large flows waste memory on the many small ones. Consequently, CMS cannot accurately capture the minority of large flows without significantly inflating overall memory usage.

To address skewness, recent works have proposed splitting large and small flows into distinct data structures, typically a key-value (KV) table for large flows combined with a compact sketch for smaller ones \cite{elasticsketch, nitrosketch, nze-sketch, bitsense}. This approach reduces collisions among different flow sizes and avoids over-allocating memory for small flows. However, a major drawback remains unresolved: when a new flow arrives, it is difficult to know immediately if it will turn into a large flow or stay small. Meanwhile, learning-based sketches attempt to predict flow size directly, hoping to bypass dynamic flow classification. LCMS \cite{lcmsketch}, for instance, trains a model to estimate whether a flow will be large, then updates either the KV table or CMS accordingly. Although this can reduce collisions when the prediction is correct, it often suffers from real-world accuracy issues and transfers poorly to dynamic network environments. Meta-sketch \cite{metasketch} takes a different route by learning the distribution of flow sizes rather than explicitly splitting flows into large and small. However, its training overhead is notably high, limiting its deployment in real-world scenarios. Other learning-based methods have explored optimizing hashing or query processing \cite{mlsketch, bertsimas2021frequency}, but they tend to share similar drawbacks -- either relying heavily on ID-size correlations or incurring significant training cost. As a result, these approaches still struggle to handle unpredictable traffic shifts while maintaining low error rates and manageable resource usage.

In this paper, we propose \alg{}, a new sketch algorithm that adapts to skewed network traffic by combining a two-tier data structure with an LLM-powered real-time flow classifier. Our key insight is that leveraging the full packet header -- beyond just the flow ID -- enables more accurate predictions of future flow sizes. By incorporating these additional header fields, \alg{} effectively infers whether a newly arriving flow is likely to become large or remain small, without relying on weak correlations between flow IDs and sizes. \alg{}’s design centers on two main techniques:

\bbb{Technique 1: Two-tier data structure.}
\alg{}’s data structure features two tiers: a \textit{heavy part} for large flows and a \textit{light part} for small flows. This design more effectively handles skewed distributions by reducing collisions that typically arise when large and small flows share the same counters. By tracking large flows in a dedicated space, \alg{} captures their sizes more accurately while preventing overflow in counters designated for small flows. Meanwhile, a compact sketch records small flows, limiting memory overhead. To determine whether a flow should be considered large, we rely on a real-time classifier. We also implement a simple \textit{lock flag} mechanism to retain historical classification results and prevent genuinely large flows from being evicted prematurely.

\bbb{Technique 2: LLM-powered flow classifier.}
\alg{} employs a \textit{flow classifier} built on a Large Language Model (LLM) adapted for network traffic. We embed each packet header (excluding specific IP addresses to avoid overfitting) into a token sequence and feed it into the model. Instead of imposing a hard threshold for labeling flows as large or small, the classifier employs a soft-label strategy: the model outputs a continuous value in \([0,1]\). Flows that are significantly larger than the threshold receive labels near 1, while those that are considerably smaller receive labels near 0. In between, flows whose predicted sizes fall near the threshold are assigned intermediate values (e.g., 0.4–0.6), capturing the inherent uncertainty and thus mitigating errors associated with borderline misclassifications, thereby improving overall accuracy of the sketch. By integrating these real-time predictions with the two-tier data structure, \alg{} can more reliably track large flows while keeping memory usage low.

We demonstrate the versatility of \alg{} through three representative network stream mining tasks: flow size query, heavy hitter query, and hierarchical heavy hitter (HHH) query. We implement \alg{} in Python and conduct extensive experiments on three real-world datasets. Experimental results show that \alg{} achieves, on average, a \textit{\(7.5\times\) improvement in accuracy} over state-of-the-art methods. All related source code is publicly available on GitHub \footnote{\url{https://github.com/LLM-Sketch/LLM-Sketch}}.

\section{Background}

\subsection{Problem Definition}

\begin{definition}
    \textit{Data Stream:} A data stream \(\mathcal{S} = \{p_1, p_2, \cdots, p_i, \cdots\}\) is a sequence of packets, where each packet \(p_i\) has a flow ID \(f\), drawn from a universe \(\mathcal{U}\). In this paper, we focus on flow measurement within a fixed time window, so the data stream is finite. Note that each packet can only be proceeded once.
\end{definition}

\begin{definition}
    \textit{Flow size query:} Given a data stream \(\mathcal{S}\), reporting the size of every flow \(f \in \mathcal{U}\), where flow size is defined as the number of packets of ID \(f\), i.e., \(n(f) = |\{i:p_i=f\}|\).
\end{definition}

\begin{definition}
    \textit{Heavy hitter query:} Given a data stream \(\mathcal{S}\) and a threshold \(T\), reporting the ID of every flow with size larger than \(T\).
\end{definition}

\begin{definition}
    \textit{Hierarchical Heavy Hitters (HHH) Query:}
    Consider a hierarchy \(H\) imposed on the flow IDs in \(\mathcal{U}\). For example, each IP address can be subdivided into multiple levels of prefixes (e.g., /8, /16, /24). Given a data stream \(\mathcal{S}\) and a threshold \(T\), an HHH query aims to find all nodes in the hierarchy whose aggregated flow size exceeds \(T\). \cite{rhhh} 
    % Formally, a node \(v\) in the hierarchy is a Hierarchical Heavy Hitter if 
    % \[
    %     \sum_{f \in \mathcal{F}(v)} n(f) > T,
    % \]
    % where \(\mathcal{F}(v)\) denotes the set of flows that map to node \(v\) (i.e., all flow IDs covered by \(v\) in the hierarchy), and \(n(f)\) is the size of flow \(f\). 
\end{definition}

\subsection{Large Language Model}

A large language model (LLM) is a neural network designed to process and generate natural language using vast amounts of training data. In recent years, it also finds success in other fields, such as biomedical research, code generation, and data analysis. Typical LLMs include BERT \cite{bert}, GPT \cite{gpt4}, and Llama \cite{llama2}. Among them, BERT (Bidirectional Encoder Representations from Transformers) introduces masked language model (MLM), enabling the model to learn context from both directions. It also leverages next sentence prediction to capture relationships between sentences. Variations of BERT include DistilBERT \cite{distilbert}, RoBERTa \cite{roberta}, and ALBERT \cite{albert}, each offering improvements in efficiency or performance.

\subsection{The Count-Min Sketch}

\begin{figure}[!ht]
    \centering  
    \includegraphics[width=0.5\linewidth]{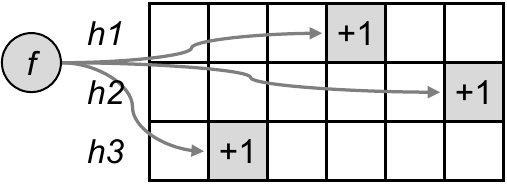}
    \caption{The Count-Min sketch.}
    \label{fig:cms}
\end{figure}

As shown in Figure \ref{fig:cms}, the data structure of the Count-Min sketch (CMS) consists of \(d\) counter array. The \(i\)-th counter array \(A_i\) consists of \(w\) counters, and is also associated with a hash function \(h_i(.) (1\leqslant h_i(.) \leqslant w)\).
When inserting a packet of flow \(f\), CMS computes the \(d\) hash functions to locate the \(d\) mapped counters, \(A_1[h_1(f)], \cdots, A_d[h_d(f)]\), and increments each mapped counter by 1.
When querying the size of flow \(f\), CMS reports the minimum value of each mapped counter as the estimated flow size, i.e., \(\min\{A_i[h_i(f)]\} (1\leqslant i \leqslant d)\).

\section{The \alg{} Algorithm}

In this section, we first propose the data structure and operations of \alg{}. Then we present how the flow classifier is designed. After that, we describe the application of \alg{}.

\subsection{Data Structure and Operations}

\begin{figure}[!ht]
    \centering  
    \includegraphics[width=\linewidth]{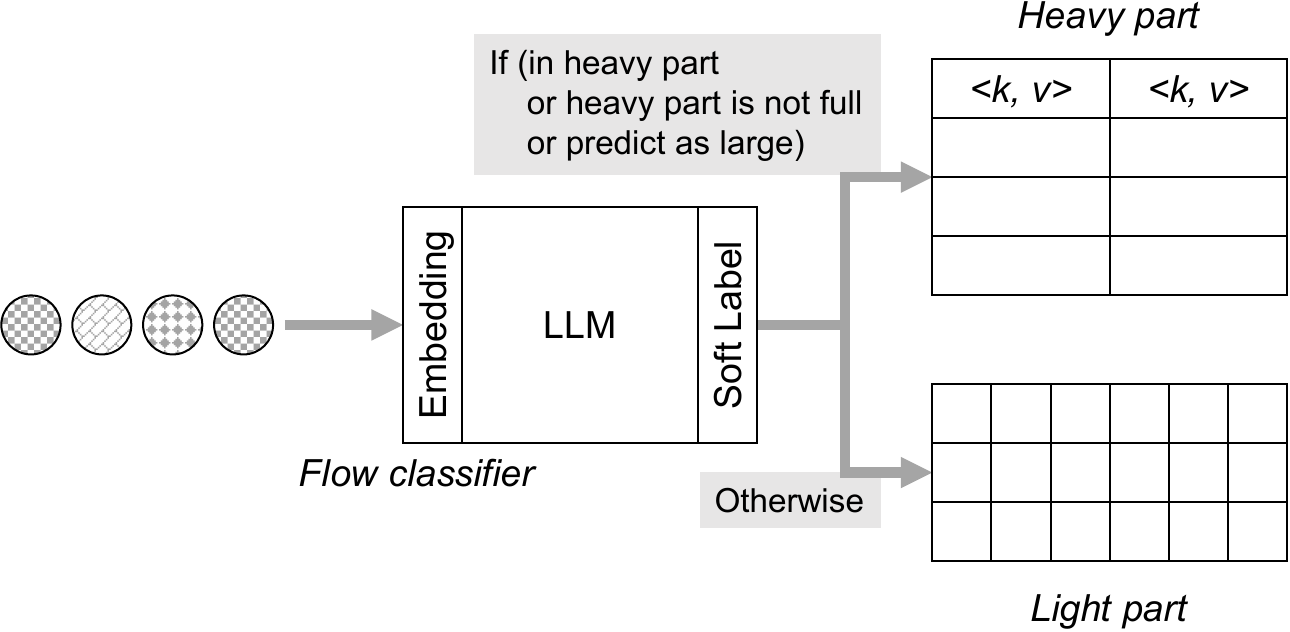}
    \caption{Workflow of \alg{}.}
    \label{fig:workflow}
\end{figure}

\bbb{Data structure:}
As shown in Figure~\ref{fig:workflow}, the data structure of \alg{} consists of three parts: a heavy part, a light part, and a flow classifier.
The heavy part is a key-value (KV) table with \(w_h\) buckets. Each bucket contains \(d_h\) cells, and each cell records a flow, including its flow ID \(f\) and flow size \(\hat{n}\). The heavy part is also associated with a hash function \(h(.) (1 \leqslant h(.) \leqslant w_h)\), which maps flows to buckets.
The light part is a CMS, which maintains the sizes of small flows using small counters (e.g., 8-bit) to save memory.
The flow classifier is a model that infers whether an incoming packet belongs to a large flow or a small flow. Its design is detailed in Section \ref{sec:alg:model}. Ideally, large flows are recorded in the heavy part, whereas the light part is used only for small flows.

\bbb{Insertion:}
When inserting a packet of flow \(f\), \alg{} locates the mapped bucket \(B[h(f)]\) using the hash function \(h\). There are three cases:

\textit{Case 1:} If \(f\) is already recorded in \(B[h(f)]\), \alg{} simply increments its flow size by 1.

\textit{Case 2:} If \(f\) is not in \(B[h(f)]\) and there is an empty cell, \alg{} inserts \((f, 1)\) into that cell.

\textit{Case 3:} If \(f\) is not in \(B[h(f)]\) and all cells in the bucket are full, \alg{} uses the flow classifier to predict whether \(f\) is a large flow or a small flow. Based on the classifier’s output, there are two sub-cases:
1) If \(f\) is a large flow, let \(f_{min}\) be the flow with the minimum flow size in \(B[h(f)]\). \alg{} evicts \(f_{min}\) from the bucket, inserts it into the light part, and then inserts \((f, 1)\) into \(B[h(f)]\).
2) If \(f\) is a small flow, \alg{} directly inserts \(f\) into the light part.

\bbb{Query:}
When querying the size of a flow \(f\), \alg{} first checks if \(f\) is in the heavy part. If so, it reports the recorded flow size; otherwise, it reports the result from the light part.

\begin{figure}[!ht]
    \centering  
    \includegraphics[width=0.8\linewidth]{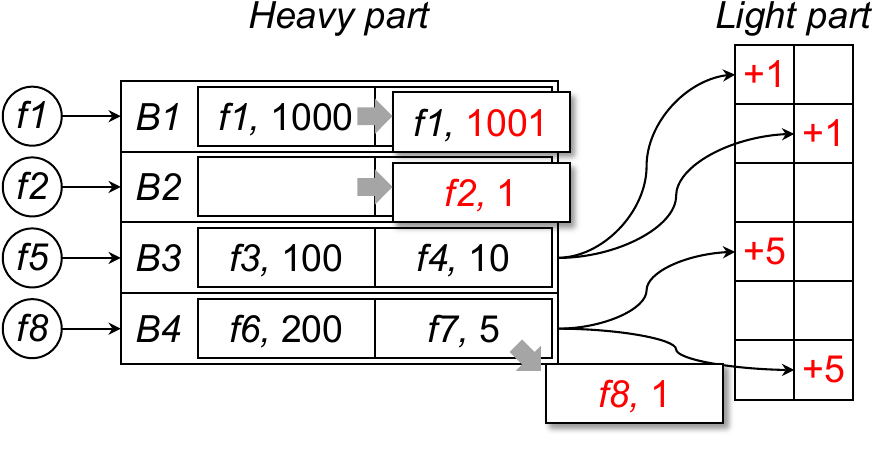}
    \caption{An example of \alg{}.}
    \label{fig:example}
\end{figure}

\textit{Example 1:} Figure~\ref{fig:example} illustrates the different cases in the insertion process of \alg{}.
When inserting a packet of flow \(f1\), \alg{} computes the hash function \(h\) to locate bucket \(B1\). Since \(f1\) is already recorded in \(B1\), \alg{} simply increments its flow size by 1.

\textit{Example 2:} When inserting a packet of \(f2\), \alg{} locates bucket \(B2\). Since \(B2\) has an empty cell, \alg{} inserts \((f2, 1)\) into that cell.

\textit{Example 3:} When inserting a packet of \(f5\), \alg{} locates bucket \(B3\). Since \(B3\) is full, \alg{} uses the flow classifier to predict that \(f5\) is a small flow and therefore inserts \((f5, 1)\) into the light part.

\textit{Example 4:} When inserting a packet of \(f8\), \alg{} locates bucket \(B4\). Since \(B4\) is full, \alg{} predicts \(f8\) as a large flow. \alg{} then evicts the flow with the minimum flow size (i.e., \(f7\)) from \(B4\), inserts \((f8, 1)\) into \(B4\), and inserts \(f7\) into the light part.

\bbb{Optimization: Large-flow Locking.}
In the insertion process described above, if a hash collision occurs, \alg{} evicts the flow with the smallest recorded size. Although this approach generally works well, it may inadvertently evict newly arrived flows that are actually large but have not yet accumulated a significant size. To address this issue, we introduce a \textit{lock flag} in each cell of the heavy part. This flag tracks how often a flow is predicted to be large, thereby reducing the likelihood of evicting flows that were previously identified as large, even if their current size is still small.

\textit{Lock flag update:}
Whenever a packet is inserted and its flow is (re)classified, we update the lock flag based on the classifier’s prediction and the flow’s recorded size.
Let \(\hat{y} \in \{0, 1\}\) be the predicted label for this packet, where \(\hat{y} = 1\) indicates large flow and \(\hat{y} = 0\) indicates small flow.
Recall that \(\hat{n}\) represents the flow's recorded size.
Then the lock flag \(L \in \{0, 1\}\) is updated as follows:
\[
L \leftarrow
\begin{cases} 
1, & \text{w.p. } \frac{L \cdot \hat{n} + \hat{y}}{\hat{n} + 1},\\
0, & \text{otherwise}.
\end{cases}
\]
This rule can be viewed as an unbiased estimator of the fraction of times the flow is predicted to be large, accumulated over its updates (See Theorem~\ref{theorem:flag}). If \(L = 1\) after the update, we treat the flow as large and therefore lock it. Otherwise, if \(L = 0\), it is more likely to be a small flow and can be safely evicted if necessary.

\textit{Eviction policy:}
When a hash collision occurs and \alg{} needs to evict a flow from a full bucket, it first checks whether any flows in that bucket have \(L = 0\). 1) If there is at least one unlocked flow (\(L = 0\)), \alg{} evicts tthe one with the smallest size among them. 2) If all flows in the bucket are locked (\(L = 1\)), \alg{} must evict the flow with the minimum size regardless of its lock flag. Although the latter scenario should be rare, it can still occur due to the classification errors or the probabilistic nature of the lock flag update.

\subsection{Model Design}
\label{sec:alg:model}

We choose to adapt a Large Language Model (LLM) as our flow classifier due to its ability to capture complex patterns in packet headers. By leveraging an LLM, we can process each packet header in a contextual manner, enabling the classifier to learn nuanced relationships that simpler models might overlook. Furthermore, the inherent flexibility of LLMs makes them well-suited for handling packet headers of varying lengths and formats.

\bbb{Embedding:}
Since the raw packet header data cannot be directly interpreted by a language model, we introduce an embedding layer that transforms the packet headers into token embeddings. Specifically, we treat the packet header as a binary string and segment it into two-byte chunks, each serving as a token for the embedding layer. This approach circumvents the need for a cumbersome, field-by-field parsing.
In practice, to prevent overfitting to specific IP addresses, we remove the source and destination IP fields before feeding the remaining header data into the model. Consequently, the classifier focuses on more generalizable features—such as transport-layer information—rather than memorizing particular endpoints in the training data.

\bbb{Objective Function:}
A straightforward strategy might be to define a hard threshold \( T \) (e.g., 64) to classify flows as large (\( \geqslant T \)) or small (\( < T \)) categories. However, directly optimizing for a strict binary cutoff can lead to the following issues:

\begin{itemize}[leftmargin=*]
    \item Flows near the threshold (e.g., those with sizes 60-70) often share similar characteristics, making a single sharp boundary somewhat arbitrary.

    \item Misclassifying flows near the threshold has relatively little impact on the overall sketch accuracy, so aggressively fitting a binary boundary can introduce unnecessary complexity.
\end{itemize}

To address these concerns, we adopt a \textit{soft-label approach} that smooths the discrete large-versus-small boundary. Concretely, we assign a label to each flow based on
\[
\text{label} = \sigma \Bigl(a \bigl((\log(n) - \log(T) \bigr)\Bigr),
\]
where \(n\) is the flow size, \(T\) is the threshold, \(\sigma(\cdot)\) is the sigmoid function, and \(a\) is a scaling parameter. This design has several advantages:

\begin{itemize}[leftmargin=*]
\item Continuity: Rather than a hard 0/1 label, flows receive labels in the continuous range \((0, 1)\), enabling a smooth transition around the threshold.

\item Reduced sensitivity: Flows that are significantly larger than \(T\) yield labels near 1, while those much smaller than \(T\) yield labels near 0. Flows in the ambiguous region around \(T\) hover near 0.5, making misclassifications less punitive.

\item Smoother optimization: Training as a regression task on these soft labels typically exhibits more stable convergence than a strict classification objective.
\end{itemize}

In practice, this soft-label mechanism helps the classifier learn a nuanced notion of flow size, rather than fixating on a single, potentially noisy threshold. Large flows (e.g., above 1,000) naturally produce labels close to 1, whereas small flows (e.g., below 5) produce labels close to 0. Flows near \(T\) fall around 0.5, diminishing the adverse impact of uncertain classifications. Consequently, the classifier achieves better overall performance and adaptability across diverse network environments.

\subsection{Application}

In this section, we describe how to apply \alg{} to 3 typical tasks: flow size query, heavy hitter query, and hierarchical heavy hitter query.

\bbb{Flow size query:}
\alg{} can be used directly to measure flow sizes.

\textit{Optimization: using fingerprints.}
Following many existing works \cite{dhs, mimosketch}, \alg{} also supports the use of fingerprints in place of full flow IDs. This is particularly beneficial when the original ID is large (e.g., the 13-byte 5-tuple).
Although fingerprints may introduce collisions, they substantially reduce memory usage. Consequently, \alg{} can achieve higher accuracy under the same memory budget compared with recording the full flow IDs.

\bbb{Heavy hitter query:}
For heavy hitter query, \alg{} maintains the same insertion procedure described earlier.
When querying heavy hitters, \alg{} simply the heavy part to find all flows whose recorded sizes exceed the given threshold. Those flows are subsequently reported as heavy hitters.

\bbb{Hierarchical heavy hitter (HHH) query:}
To support HHH query, \alg{} replaces the light part with a CocoSketch \cite{cocosketch}. The insertion process remains primarily unchanged, except that flows which would originally be inserted into CMS are now inserted into CocoSketch. When performing an HHH query, \alg{} first merges the heavy and light parts into a single key-value table and then obtains HHH using the aggregation approach proposed by CocoSketch.

% \bbb{DDoS victim query:}
% %
% \alg{} works in conjunction with a Bloom filter (BF) \cite{bloomfilter} for DDoS victim query, where BF is used for deduplication and \alg{} is used for counting. When a packet with ID \(f = \langle src, dst\rangle\) is inserted, we first query BF to check whether \(f\) has already been inserted. If BF returns true, i.e. \(f\) is already recorded, then no action is taken. If the BF returns false, we insert \(f\) into BF, and \((dst, 1)\) into \alg{}.
% %
% When querying for DDoS, \alg{} scans the heavy part, and reports the set of flows whose corresponding values are greater than the threshold \(T_{DDoS}\).

\section{Theoretical Analysis}

\subsection{Accuracy and Error Bound}

We make the following assumptions in our analysis.

\begin{itemize}[leftmargin=*]
    \item \textbf{Assumption 1 (Classification Consistency):} The predicted label of a flow does not change from large to small (or vice versa) during its lifetime.

    \item \textbf{Assumption 2 (Sufficient Heavy Part):} The heavy part is large enough so that any large flow correctly classified as large is never evicted once inserted.
\end{itemize}

\begin{theorem}
\label{theorem:full_accurate}
The probability that a large flow is fully accurate (i.e., tracked with zero error) in \alg{} is
\[
  P_{\mathrm{LLMS}} = A + (1 - A) \cdot P_{\mathrm{CMS}} \bigl(w_{light}, d_{light}, N_{light}\bigr),
\]
where \(A\) is the classifier's accuracy for large flows, i.e., the probability that a large flow is correctly identified as large.
\(w_{light}\), \(d_{light}\) is the width and depth of the light part (CMS), and \(N_{light}\) is the number of flows that end up in the light part.
\(P_{\mathrm{CMS}}(w, d, N)\) is the probability that a single flow is fully accurate in a CMS of width \(w\) and depth \(d\), when there are \(N\) total flows in that sketch. One common formula, based on a Poisson approximation of collisions, is:
\[
  P_{\mathrm{CMS}}(w, d, N) = 1 - \bigl(1 - e^{-(N-1)/w}\bigr)^{d}.
\]
\end{theorem}

\begin{proof}
Consider a large flow \(f\):

\bbb{Case 1: Correct Classification (probability \(A\)).}
Under Assumption 2, once inserted into the heavy part, \(f\) is never evicted. Consequently, \(f\) is fully accurate, so the probability in this scenario is \(1\).

\bbb{Case 2: Misclassification (probability \(1 - A\)).}
If \(f\) is misclassified as small, it goes into the light part.
The probability that \(f\) is fully accurate in CMS is \(P_{\mathrm{CMS}}(w_{light}, d_{light}, N_{light})\).

By the law of total probability, we sum over these two disjoint cases:
\begin{align*}
    P_{\mathrm{LLMS}}
    &= A \times 1 + (1 - A) \times P_{\mathrm{CMS}}(w_{light}, d_{light}, N_{light}) \\
    &= A + (1 - A) \cdot P_{\mathrm{CMS}} \bigl(w_{light}, d_{light}, N_{light}\bigr).
\end{align*}
\end{proof}

\begin{theorem}
\label{theorem:error_bound}
Let \(\hat{n(f)}\) and \(n(f)\) be the estimated and actual size of a flow \(f\).
Let \(\|\mathbf{f_{light}}\|_1\) be the total number of packets sent to the light part. We have
    \[\|\mathbf{f_{light}}\|_1 \leqslant \|\mathbf{f}\|_1 - A \times N_{large} \times T,
\]
where \(\|\mathbf{f}\|_1\) is the total number of packets.
Let \(\delta = (1 - A) e^{-d_{light}}\), \(\epsilon = \frac{e}{w_{light}}\).
With probability at least \(1 - \delta\), 
\[
    \hat{n(f)} \leqslant n(f) + \epsilon \|\mathbf{f_{light}}\|_1.
\]
\end{theorem}

\begin{proof}
We split into two cases according to whether the flow $f$ is correctly or incorrectly classified:

\textbf{Case 1: Correct Classification (probability \(A\)).}
If $f$ is a large flow and the classifier labels it as large, $f$ is inserted into the heavy part. 
Under our assumptions, once in the heavy part, $f$ is fully accurate. 
Thus, in this scenario, we trivially have
\[
  \hat{n(f)} = n(f) \leqslant n(f) + \epsilon \|\mathbf{f_{light}}\|_1.
\]

\textbf{Case 2: Misclassification (probability $1 - A$).}
If $f$ is a large flow but is incorrectly labeled as small, it goes into the light part. 
Standard CMS analysis shows that with probability $1 - e^{-d_{light}}$,
\[
  \hat{n(f)} \leqslant n(f) + \epsilon \|\mathbf{f_{light}}\|_1.
\]
Equivalently,
\[
  P\Bigl(\hat{n(f)} > n(f) + \epsilon \|\mathbf{f_{light}}\|_1 \bigm| \text{misclassified}\Bigr)
  \leqslant e^{-d_{light}}.
\]

Hence, unconditionally, the probability that \(f\) ends up in CMS and is over-counted by more than \(\epsilon \|\mathbf{f_{light}}\|_1\) is at most
\[
  (1 - A) e^{-d_{light}}.
\]

Therefore, with probability at least \(1 - (1 - A) e^{-d_{light}}\),
\[
    \hat{n(f)} \leqslant n(f) + \epsilon \|\mathbf{f_{light}}\|_1.
\]

Combining both cases, we set
\[
  \delta = (1 - A) e^{-d_{light}},
\]
which gives
\[
    P\Bigl(\hat{n(f)} \leqslant n(f) + \epsilon \|\mathbf{f_{light}}\|_1\Bigr) \geqslant 1 - \delta.
\]

Finally, observe that
\[
    \|\mathbf{f_{light}}\|_1 \leqslant \|\mathbf{f}\|_1 - A \times N_{large} \times T
\]
because at least $A\times N_{large}$ large flows are correctly classified and stored in the heavy part, each contributing at least $T$ packets. 
Hence, the light part contains at most $\|\mathbf{f}\|_1 - A\times N_{large} \times T$ packets in total, ensuring that 
\[
    \hat{n(f)} \leqslant n(f) + \epsilon \Bigl(\|\mathbf{f}\|_1 - A\times N_{large} \times T\Bigr).
\]

\end{proof}

\subsection{Lock Flag Estimation}

\begin{theorem}
\label{theorem:flag}
The lock flag \(L\) is an unbiased estimator of the flow’s predicted labels.
\end{theorem}

\begin{proof}
Let \(\hat{n}_t\) be the flow size for a given flow up to the \(t\)-th packet, and \(\hat{y}_t\in\{0,1\}\) be the classifier’s predicted label at the \(t\)-th insertion. Let \(L_t\) be the value of the lock flag after the \(t\)-th insertion. The lock flag is updated as follows:

\[
L_{t+1} = \begin{cases}
    1, & \text{w.p.}\, \frac{L_t \cdot \hat{n}_{t} + \hat{y}_{t+1}}{\hat{n}_t + 1} \\
    0, & \text{otherwise}
\end{cases}
\]

We aim to show that
\[
    \mathbb{E}[L_t] = \frac{1}{t} \sum_{i=1}^t \mathbb{E}[\hat{y}_i]
\]

\textbf{Base Case (\(t=1\)).}
When \(t=1\), the flow has been inserted exactly once. Thus,
\[
L_1 = \begin{cases}
    1, & \text{w.p.}\, \hat{y}_1 \\
    0, & \text{otherwise}
\end{cases}
\]

Hence,
\[
\mathbb{E}[L_1] = 1 \times \hat{y}_1 + 0 \times (1 - \hat{y}_1) = \hat{y}_1
\]

Since the expectation of the predicted labels for just one insertion is \(\hat{y}_1\), we have
\[
    \mathbb{E}[L_1] = \frac{1}{1}\sum_{i=1}^1 \mathbb{E}[\hat{y}_i] = \mathbb{E}[\hat{y}_1]
\]

Thus, the base case holds.

\textbf{Inductive Step.}
Assume that after \(t\) insertions,
\[
\mathbb{E}[L_t] = \frac{1}{t}\sum_{i=1}^t \mathbb{E}[\hat{y}_i]
\]

We want to show that the statement also holds for \(t+1\), i.e.,
\[
\mathbb{E}[L_{t+1}] = \frac{1}{t+1}\sum_{i=1}^{t+1} \mathbb{E}[\hat{y}_i]
\]

By the law of total expectation and the update rule for \(L\), we have
\begin{align*}
\mathbb{E}[L_{t+1}]
    &= \mathbb{E} \Bigl[\mathbb{E}\bigl[L_{t+1} \big\vert L_t, \hat{n}_t, \hat{y}_{t+1} \bigr]\Bigr] \\
    &= \mathbb{E} \Bigl[\frac{L_t \cdot \hat{n}_t + \hat{y}_{t+1}}{\hat{n}_t + 1}\Bigr]
\end{align*}

Since \(\hat{n}_t = t\) at the \((t+1)\)-th insertion, we have
\begin{align*}
\mathbb{E}[L_{t+1}]
    &= \mathbb{E}\Bigl[\frac{L_t \cdot t + \hat{y}_{t+1}}{t + 1} \Bigr]\\
    &= \frac{1}{t+1} \mathbb{E}\bigl[L_t \cdot t + \hat{y}_{t+1} \bigr]\\
    &= \frac{1}{t+1} \Bigl[t \cdot \mathbb{E}[L_t] + \mathbb{E}[\hat{y}_{t+1}] \Bigr]
\end{align*}

Using the inductive hypothesis
\(\mathbb{E}[L_t] = \frac{1}{t}\sum_{i=1}^t \mathbb{E}[\hat{y}_i]\),
we have
\begin{align*}
\mathbb{E}[L_{t+1}]
    &= \frac{1}{t+1}\Bigl[t \cdot \frac{1}{t}\sum_{i=1}^t \mathbb{E}[\hat{y}_i] + \mathbb{E}[\hat{y}_{t+1}]\Bigr]\\
    % &= \frac{1}{t+1}\Bigl[\sum_{i=1}^t \mathbb{E}[\hat{y}_i] + \mathbb{E}[\hat{y}_{t+1}]\Bigr]\\
    &= \frac{1}{t+1} \sum_{i=1}^{t+1} \mathbb{E}[\hat{y}_i].
\end{align*}

This completes the inductive step.

\textbf{Conclusion.}
By mathematical induction, we have
\[
\mathbb{E}[L_t] = \frac{1}{t}\sum_{i=1}^t \mathbb{E}[\hat{y}_i]
\]
for all \(t\). Therefore, \(L_t\) is an unbiased estimator of the average predicted label up to the \(t\)-th insertion.
\end{proof}

\section{Experimental Results}

\subsection{Experimental Setup}

\textbf{Computation platform and implementation:}
We conduct all experiments on a GPU server (Intel i9-10980XE) equipped with a NVIDIA-4090 GPU (with 24GB memory).
We implement \alg{} (Ours), Learned CM sketch (LCMS) \cite{lcmsketch}, meta-sketch (MS) \cite{metasketch}, ElasticSketch (ES) \cite{elasticsketch}, and CocoSketch (CocoS) \cite{cocosketch} in Python.
For heavy hitter query, we set the threshold \(T = 0.01\% \cdot \|\mathbf{f_l}\|_1\) as many papers do.
For \alg{}, we use Roberta \cite{roberta} as the base model, and set the soft label to.
\[
    \text{label} = \sigma \Bigl(2.298((\log_2(n) - \log_2(64))\Bigr)
\]
Hence, for flows whose sizes exceed 256, the label is above 0.99; while for flows whose sizes are below 16, the label is below 0.01.
We apply LoRA (Low-Rank Adaptation) \cite{lora} to fine-tune our model. We limit the training process to 1 epoch due to fast convergence.
For other algorithms, we set their parameters as the original paper recommended.

\bbb{Datasets:} The datasets used for the evaluation are listed as follows.
\begin{itemize}[leftmargin=*]
    \item \textit{CAIDA dataset} \footnote{\url{https://www.caida.org/catalog/datasets/passive_dataset}} is a passive traces dataset collected from high-speed monitors on a commercial backbone link.
    We split the dataset into sub-datasets with time window of 5s.
    Each sub-dataset consists of approximate 190k flows, 2.1M packets.
    We use 12 adjacent sub-dataset as the training set, and 1 sub-dataset as the test set.
    \item \textit{MAWI dataset} \footnote{\url{https://mawi.wide.ad.jp/mawi}} is a traffic dataset maintained by the MAWI Working Group of the WIDE Project, collected at the transit link of WIDE to the upstream ISP.
    We split the dataset into sub-datasets with time window of 5s.
    Each sub-dataset consists of approximate 670k flows, 1.2M packets.
    We use 12 adjacent sub-dataset as the training set, and 1 sub-dataset as the test set.
    \item \textit{IMC DC dataset} \footnote{\url{https://pages.cs.wisc.edu/~tbenson/IMC10_Data.html}} contains data from the data centers studied in \cite{imcdc}.
    The dataset consists of multible sub-datasets, each consists of approximate 63k flows, 0.9M packets.
    We use 10 sub-dataset as the training set, and 1 sub-dataset as the test set.
\end{itemize}

\bbb{Metrics:}
The metrics used for the evaluation are listed as follows.
\begin{itemize}[leftmargin=*]
    \item \textit{Average Relative Error (ARE):}
    \(\frac{1}{|\mathcal{U}|} \sum_{f\in\mathcal{U}} \frac{|n(f)-\hat{n(f)}|}{n(f)}\),
    where \(\mathcal{U}\) is the universe, \(n(f)\) and \(\hat{n(f)}\) are the actual and estimated flow size of flow \(f\), respectively.
    \item \textit{Average Absolute Error (AAE):}
    \(\frac{1}{|\mathcal{U}|} \sum_{f\in\mathcal{U}} |n(f)-\hat{n(f)}|\).
    \item \textit{F1 Score:} \(\frac{2\cdot PR\cdot RR}{PR+RR}\), where PR (Precision Rate) is the ratio of true positive flows to all reported flows, and RR (Recall Rate) is the ratio of true positive flows to all actual flows.
    %
    % \item \textit{Weighted Mean Relative Error (WMRE):} $\frac{\sum_{i=1}^{z}|m_i-\hat{m_i}|}{\sum_{i=1}^{z} \left(\frac{m_i+\hat{m_i}}{2}\right)}$, where $m_i$ and $\hat{m_i}$ are the true and estimated numbers of the items of frequency $i$, respectively, and $z$ is the maximum frequency.
    %
    % \item \textit{Related Error (RE):} \(\frac{|True-Est|}{True}\), where \(True\) and \(Est\) are the true and estimated values, respectively.
    %
    % \item \textit{Throughput:} The number of packets per second, in million packets per second (Mpps).
\end{itemize}

\subsection{Experiment on Parameter Settings}

\begin{figure}[t!]
    \centering
    \subfigure[ARE.]{
    \includegraphics[width=0.225\textwidth]{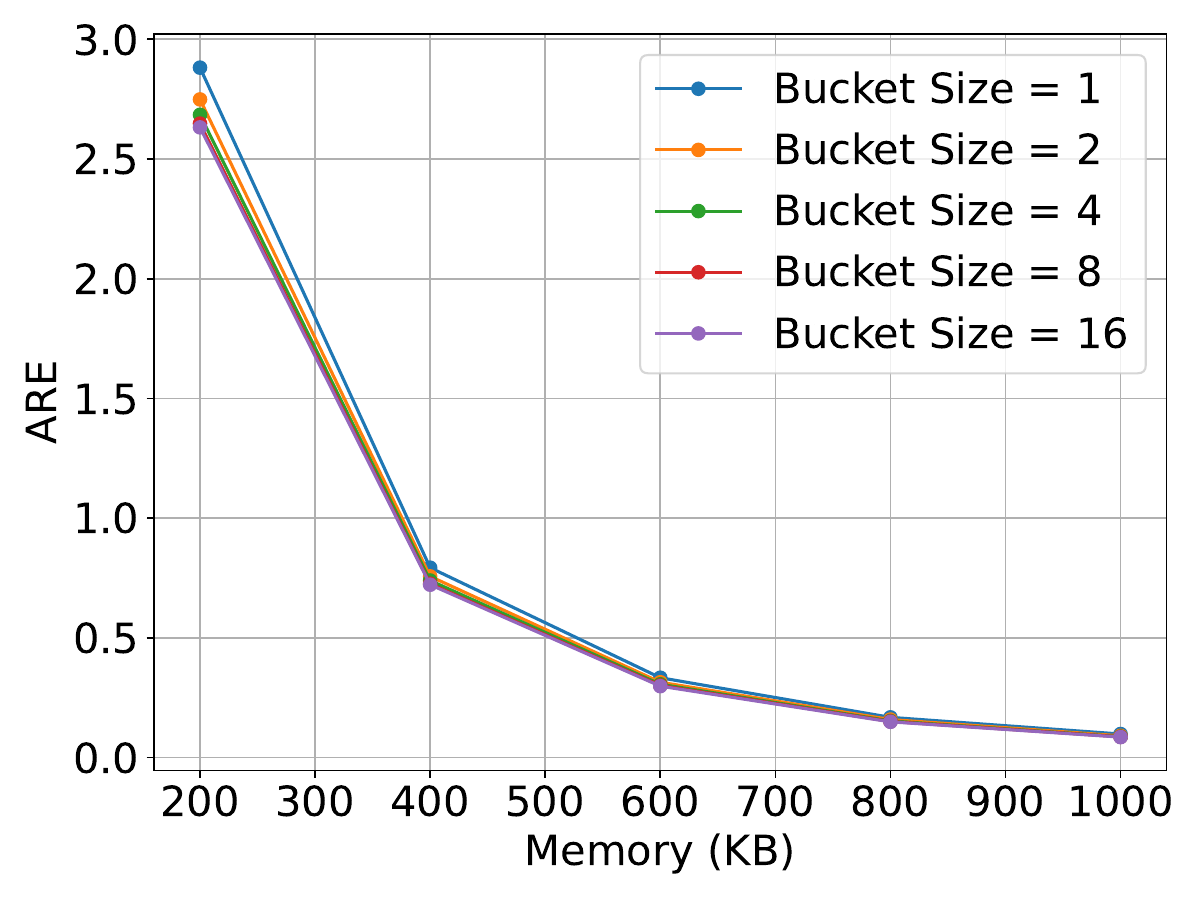}
    \label{eva:bucket_size:are}
    }
    \subfigure[AAE.]{
    \includegraphics[width=0.225\textwidth]{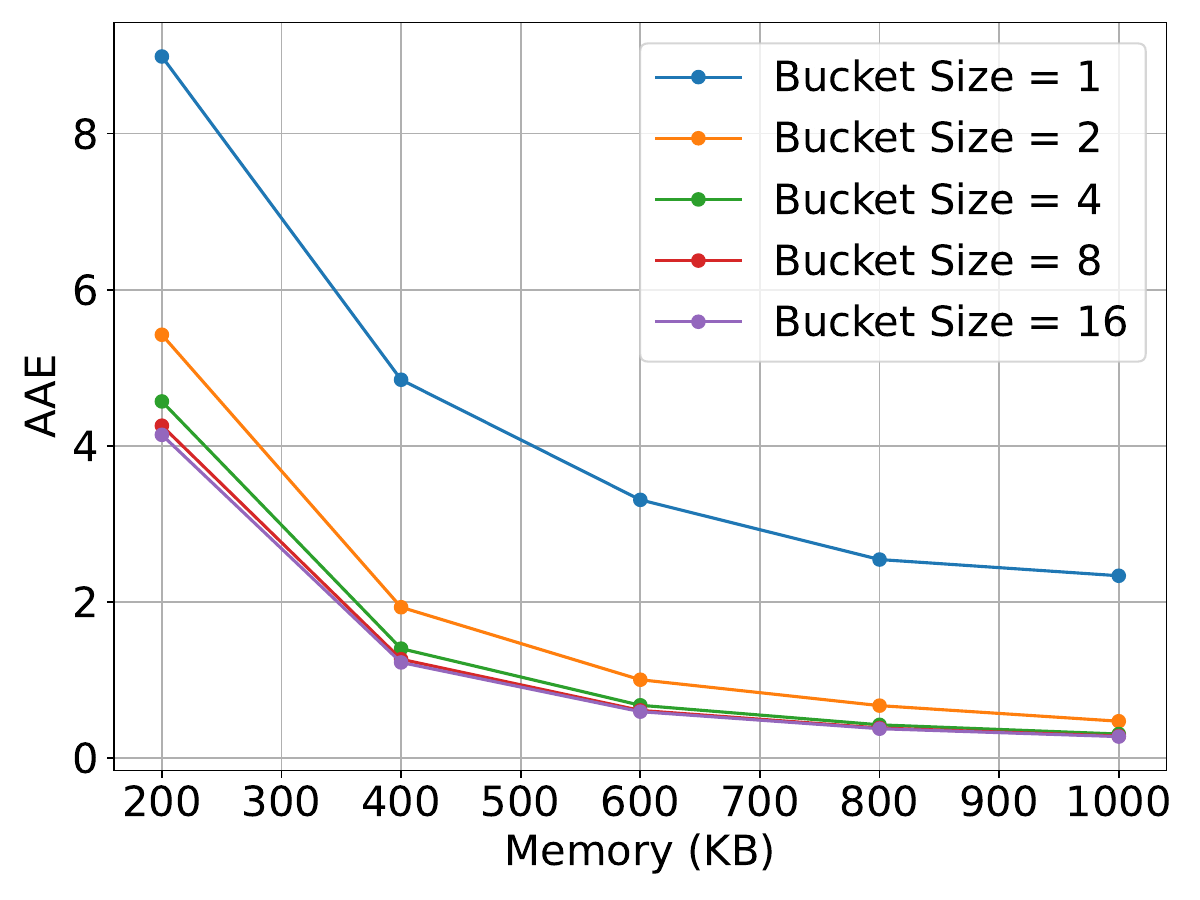}
    \label{eva:bucket_size:aae}
    }
    \caption{Accuracy vs. \# bucket size.}
    \label{eva:bucket_size}
\end{figure}

\begin{figure}[t!]
    \centering
    \subfigure[ARE.]{
    \includegraphics[width=0.225\textwidth]{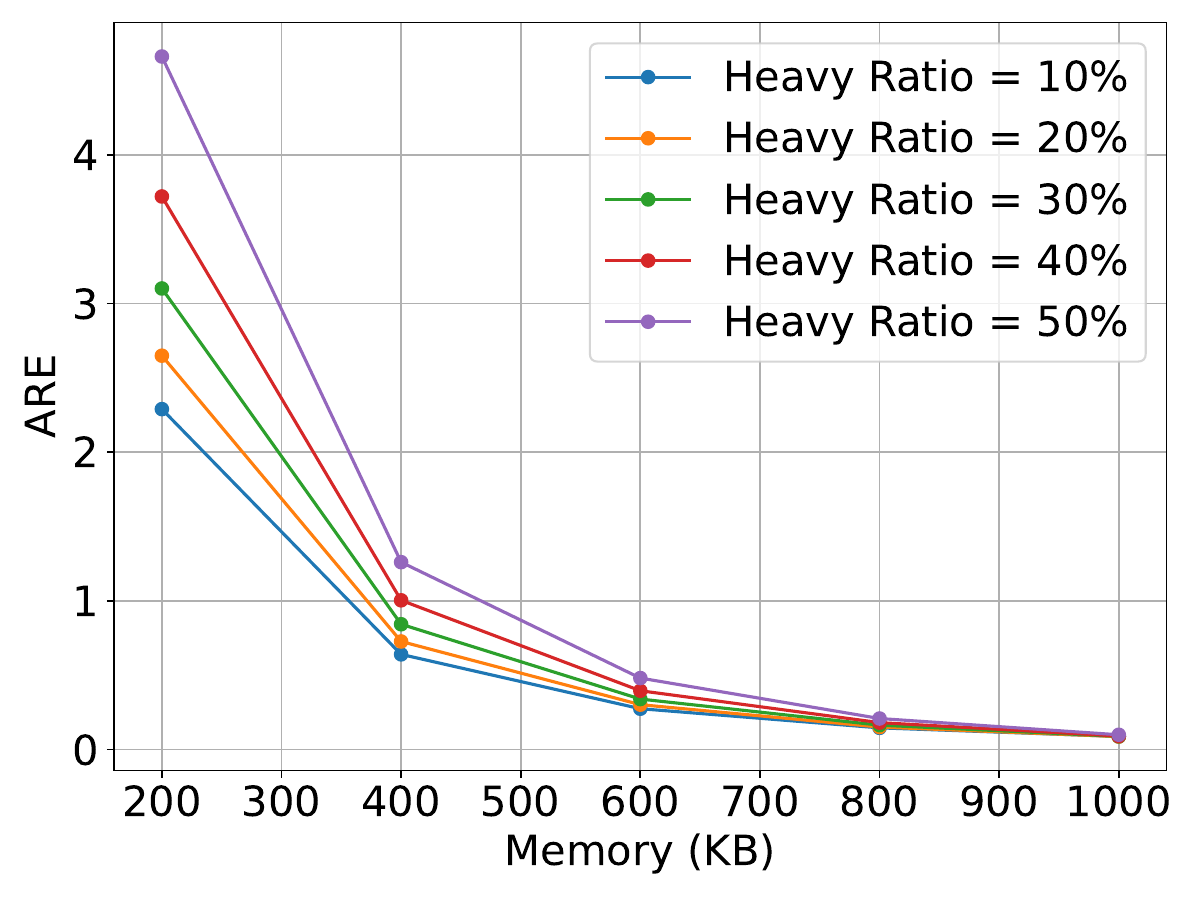}
    \label{eva:heavy_ratio:are}
    }
    \subfigure[AAE.]{
    \includegraphics[width=0.225\textwidth]{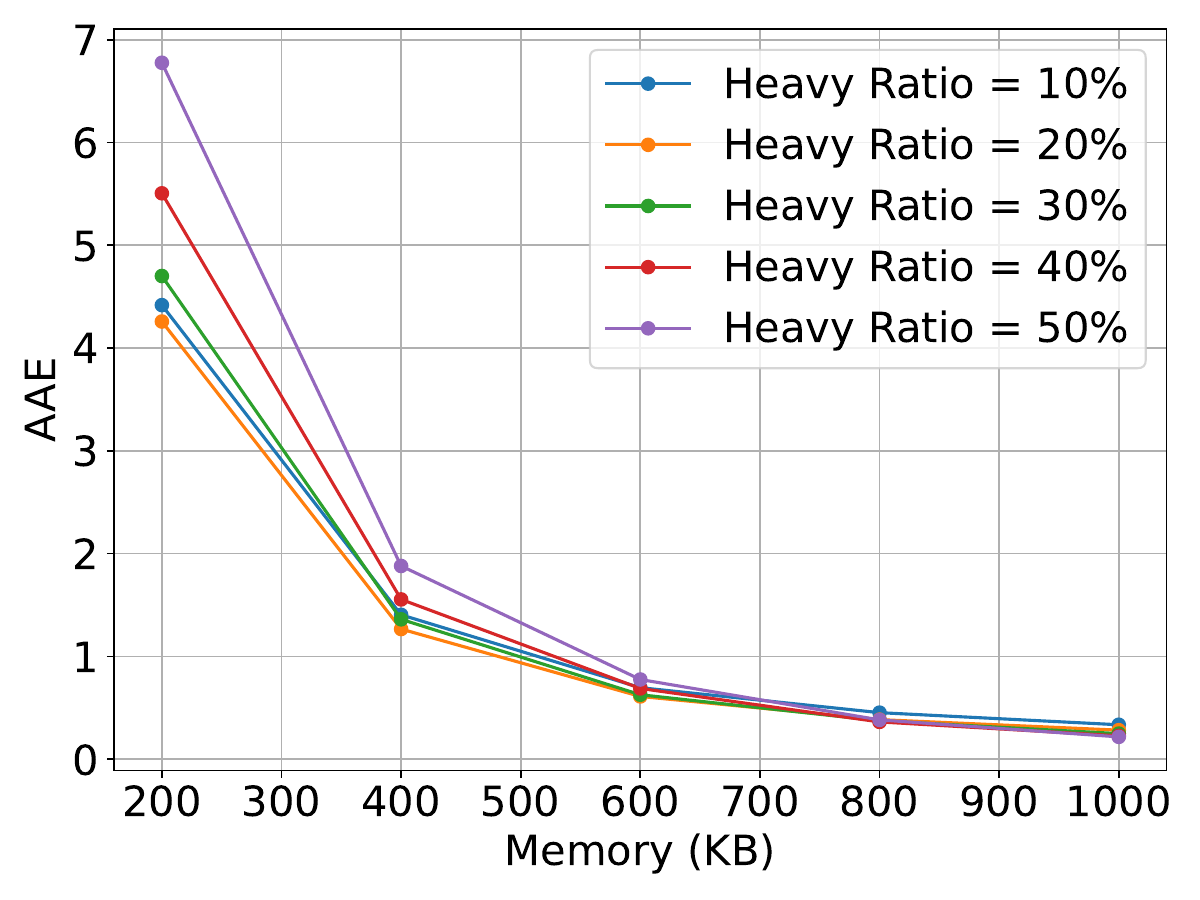}
    \label{eva:heavy_ratio:aae}
    }
    \caption{Accuracy vs. heavy ratio.}
    \label{eva:heavy_ratio}
\end{figure}

\begin{figure}[t!]
    \centering
    \subfigure[ARE.]{
    \includegraphics[width=0.225\textwidth]{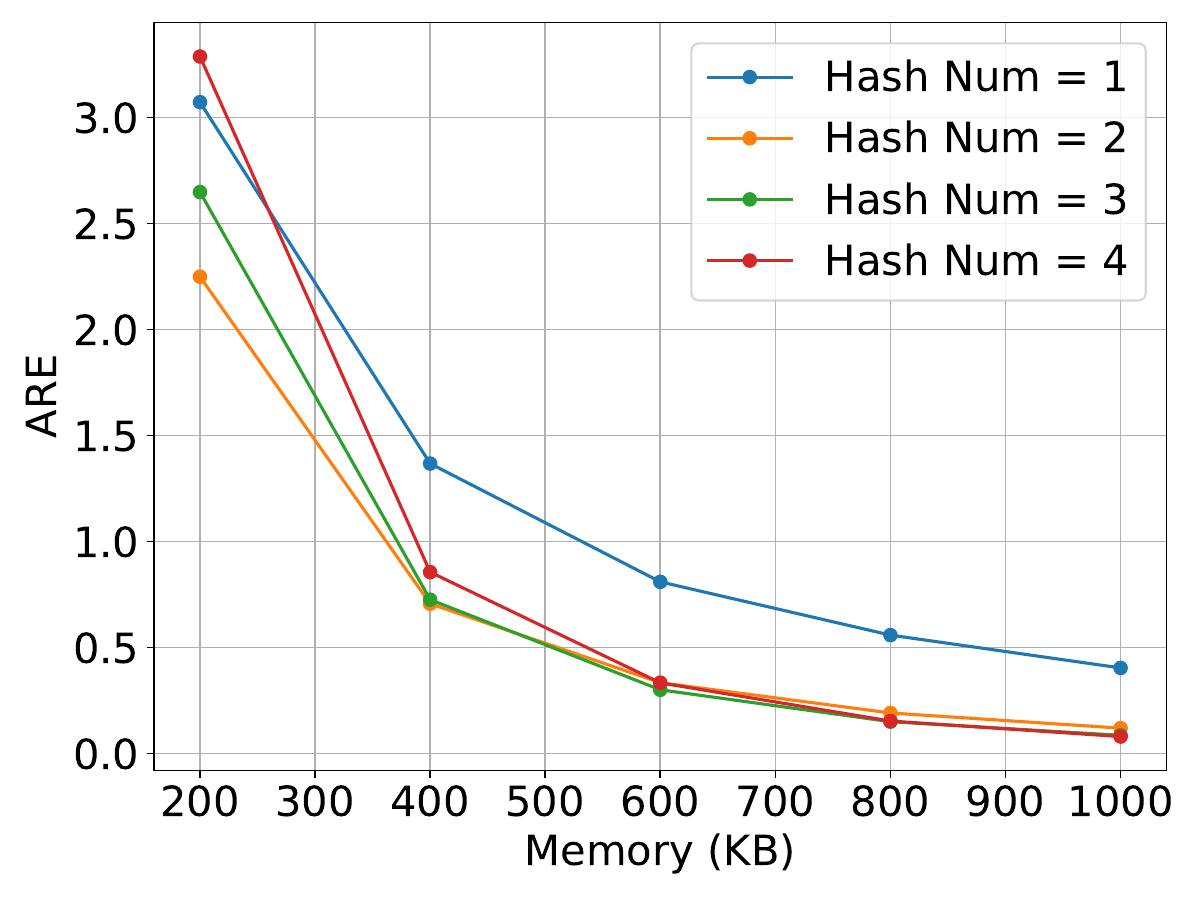}
    \label{eva:hash_num:are}
    }
    \subfigure[AAE.]{
    \includegraphics[width=0.225\textwidth]{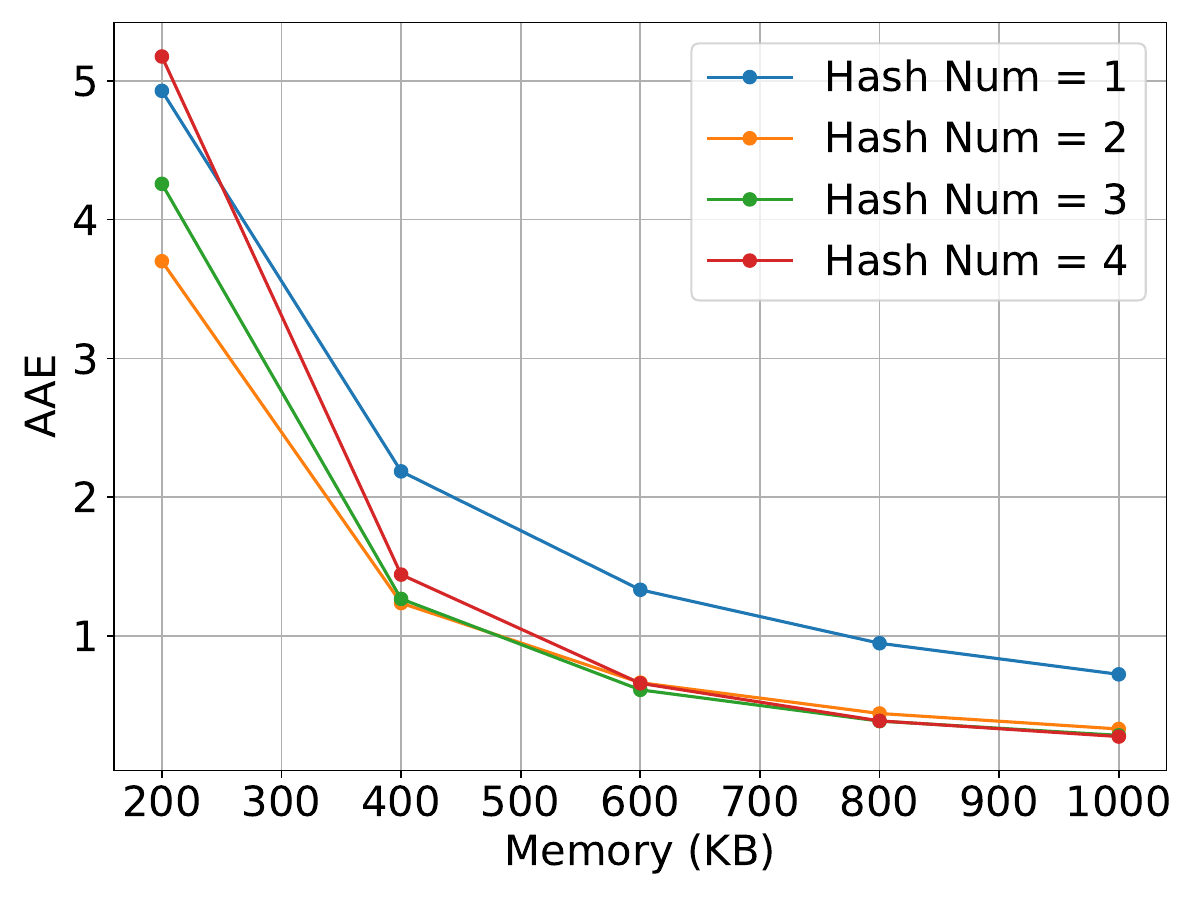}
    \label{eva:hash_num:aae}
    }
    \caption{Accuracy vs. \# hash functions.}
    \label{eva:hash_num}
\end{figure}

\bbb{\# bucket size (Figure~\ref{eva:bucket_size}):}
We vary the bucket size used in the heavy part of \alg{}. We find that when the bucket size is 1, the error is highest because two large flows may collide in the same bucket. As the bucket size increases, accuracy improves. When the bucket size is at least 2, even if collisions occur, the two large flows can still be recorded in different cells within the same bucket, thus avoiding collision-induced errors. Once the bucket size reaches 8, further increases lead to only marginal improvements.
Therefore, we set the bucket size in the heavy part to 8.

\bbb{Heavy ratio (Figure~\ref{eva:heavy_ratio}):}
We adjust the proportion of total memory allocated to the heavy part (referred to as the heavy ratio) and measure the accuracy. We find that a heavy ratio of 10\% consistently yields the lowest ARE, while the heavy ratio that achieves the lowest AAE varies with the total memory size. This is because increasing the heavy ratio improves the accuracy of large flows but lowers the accuracy of small flows.
When the total memory is small, the accuracy of small flows has a greater impact on overall accuracy, so a smaller heavy ratio yields better results. Conversely, when the total memory is large, the accuracy of large flows has a bigger influence, and AAE becomes more sensitive to large-flow accuracy, so a larger heavy ratio leads to better overall performance.
Therefore, for flow size query, we set the heavy ratio to 20\% as a balance between the accuracy of large flows and small flows.
Note that for heavy hitter query, we only use the heavy part, because heavy hitter query focuses solely on large-flow accuracy.

\bbb{\# hash functions (Figure~\ref{eva:hash_num}):}
We vary the number of hash functions used in the light part of \alg{} and examine its accuracy. We observe that when there is only 1 hash function, the error is highest due to the lack of multi-hash error reduction. When the number of hash functions is at least two, the best-performing number of hashes depends on the memory size. Nevertheless, with 3 hash functions, \alg{} consistently achieves near-optimal accuracy.
Therefore, we set the number of hash functions in the light part to 3.

\subsection{End-to-end Performance}

\begin{figure}[t!]
    \centering
    \subfigure[ARE.]{
    \includegraphics[width=0.225\textwidth]{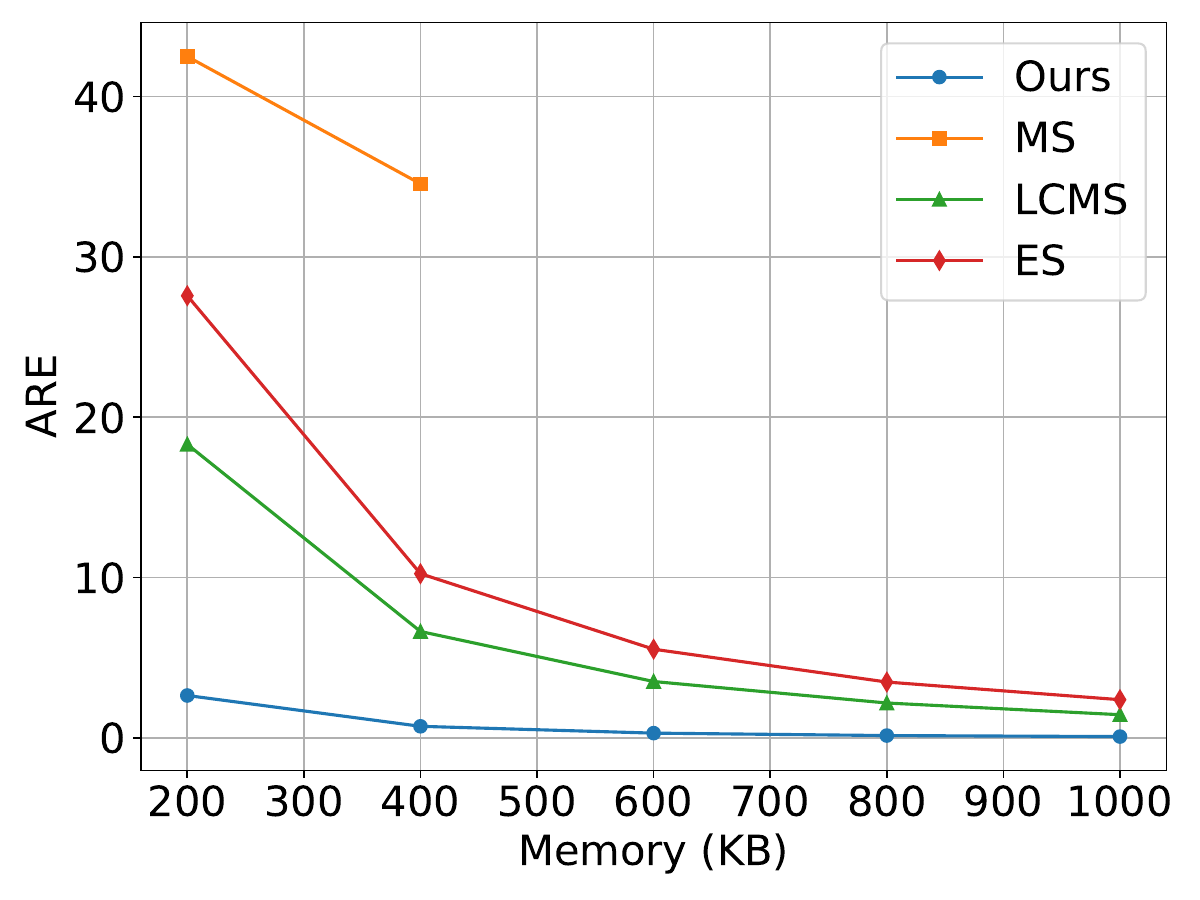}
    \label{eva:fs:are}
    }
    \subfigure[AAE.]{
    \includegraphics[width=0.225\textwidth]{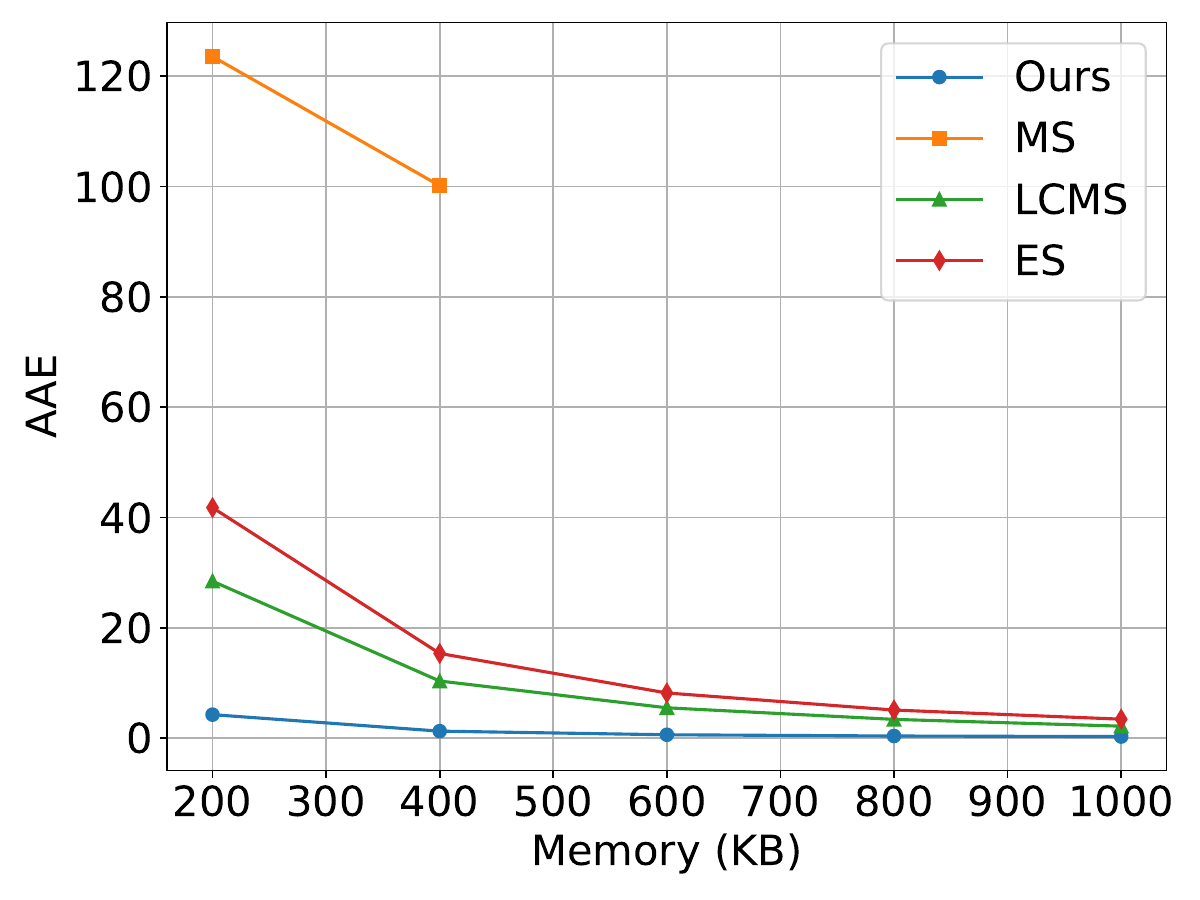}
    \label{eva:fs:aae}
    }
    \caption{Accuracy of flow size query on CAIDA dataset.}
    \label{eva:fs}
\end{figure}

\begin{figure}[t!]
    \centering
    \subfigure[F1 score.]{
    \includegraphics[width=0.225\textwidth]{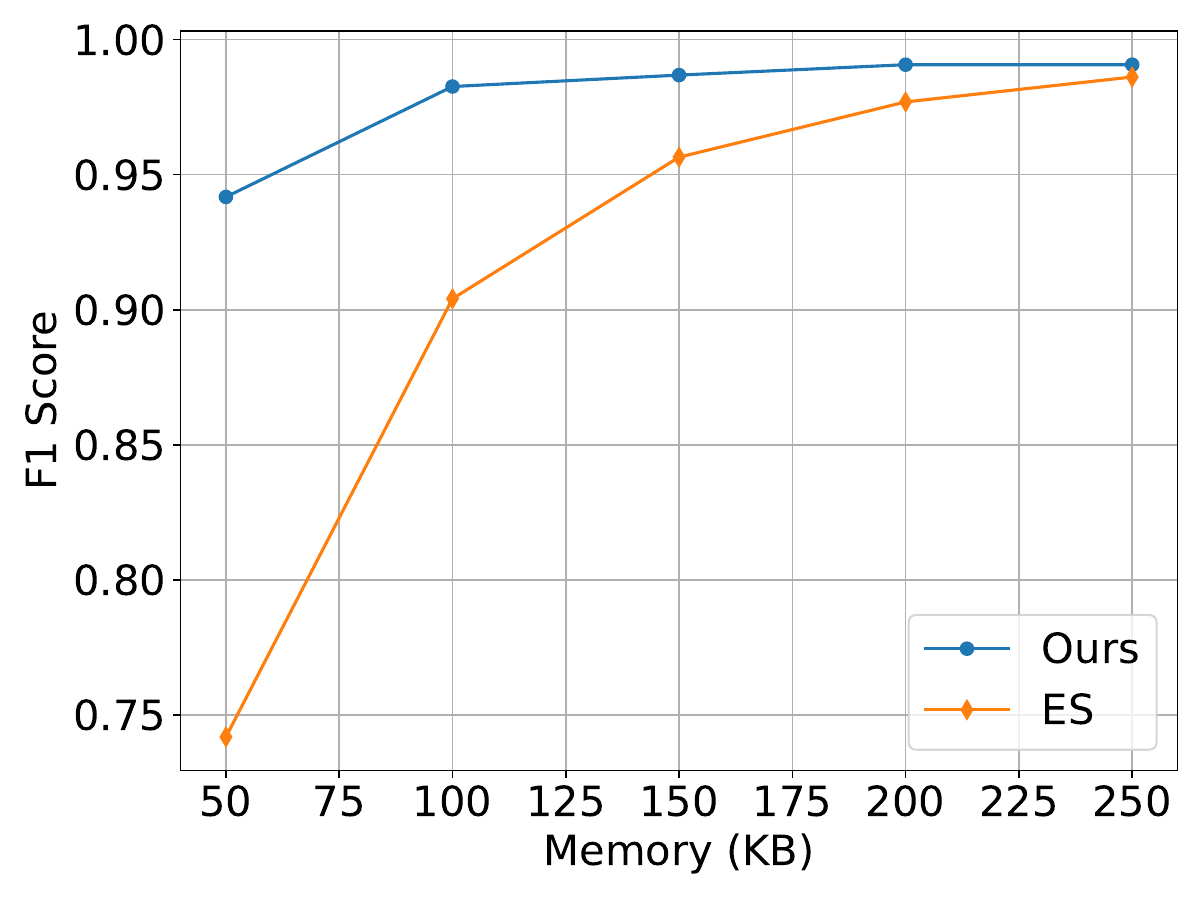}
    \label{eva:hh:f1}
    }
    \subfigure[ARE.]{
    \includegraphics[width=0.225\textwidth]{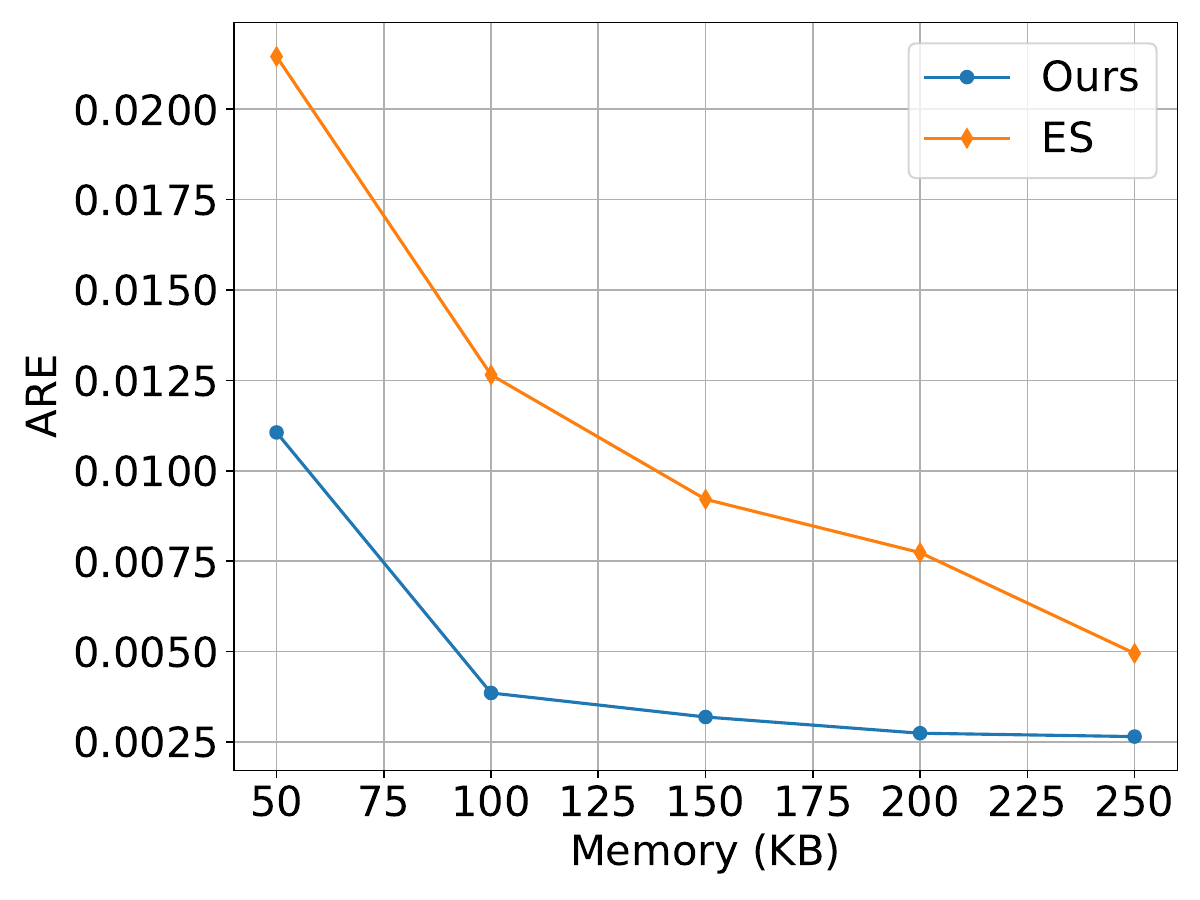}
    \label{eva:hh:are}
    }
    \caption{Accuracy of heavy hitter query on CAIDA dataset.}
    \label{eva:hh}
\end{figure}

\begin{figure}[t!]
    \centering
    \subfigure[F1 score.]{
    \includegraphics[width=0.225\textwidth]{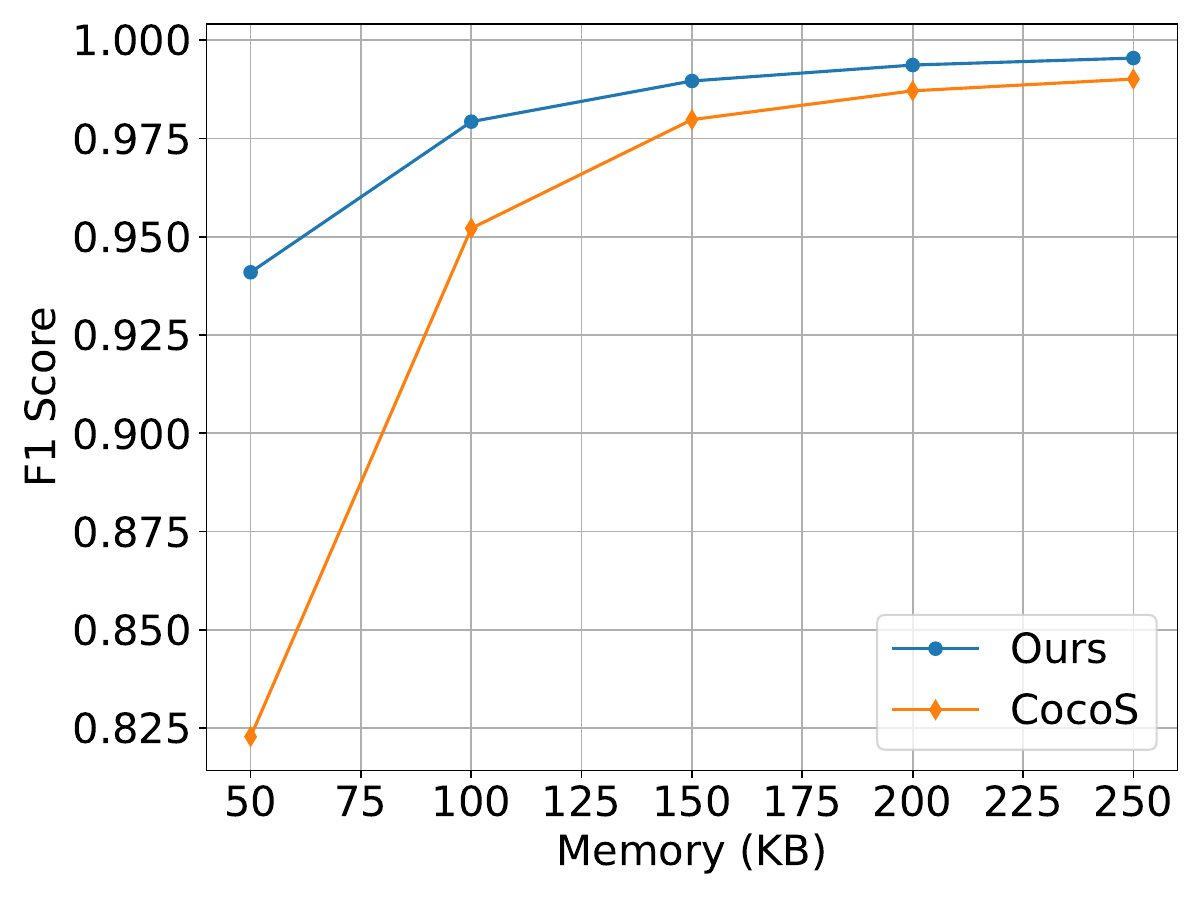}
    \label{eva:hhh:f1}
    }
    \subfigure[ARE.]{
    \includegraphics[width=0.225\textwidth]{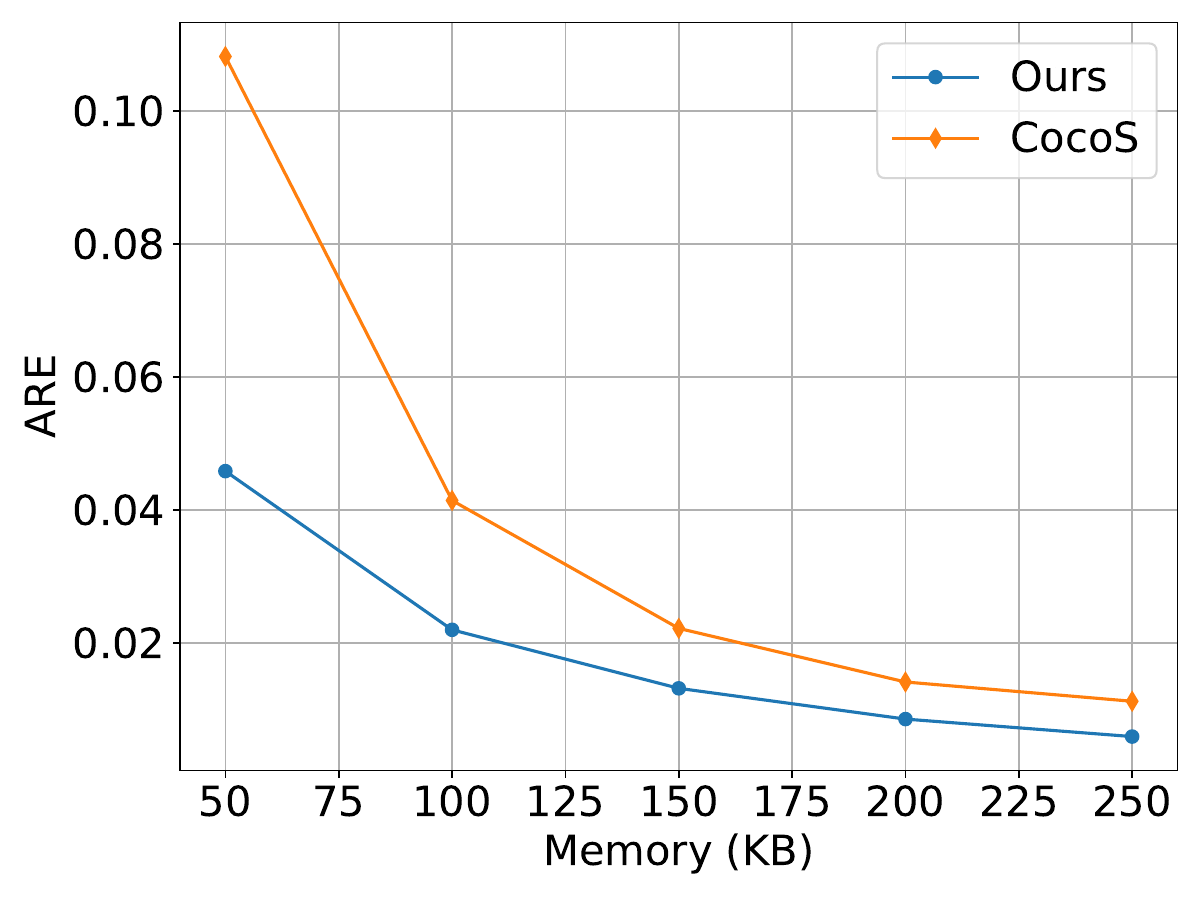}
    \label{eva:hhh:are}
    }
    \caption{Accuracy of HHH query on CAIDA dataset.}
    \label{eva:hhh}
\end{figure}

\begin{figure*}[t!]
    \centering
    \subfigure[ARE on MAWI.]{
    \includegraphics[width=0.225\textwidth]{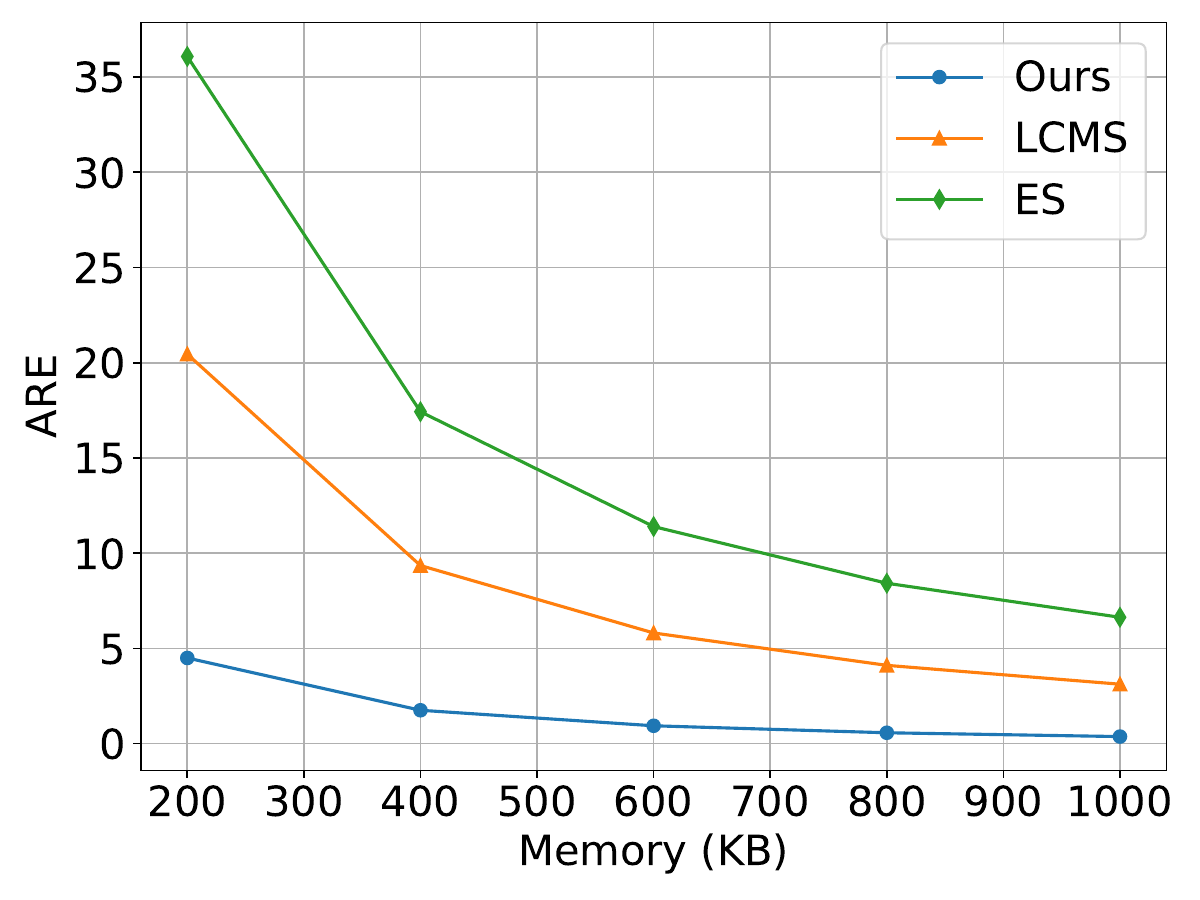}
    \label{eva:fs_mawi:are}
    }
    \subfigure[AAE on MAWI.]{
    \includegraphics[width=0.225\textwidth]{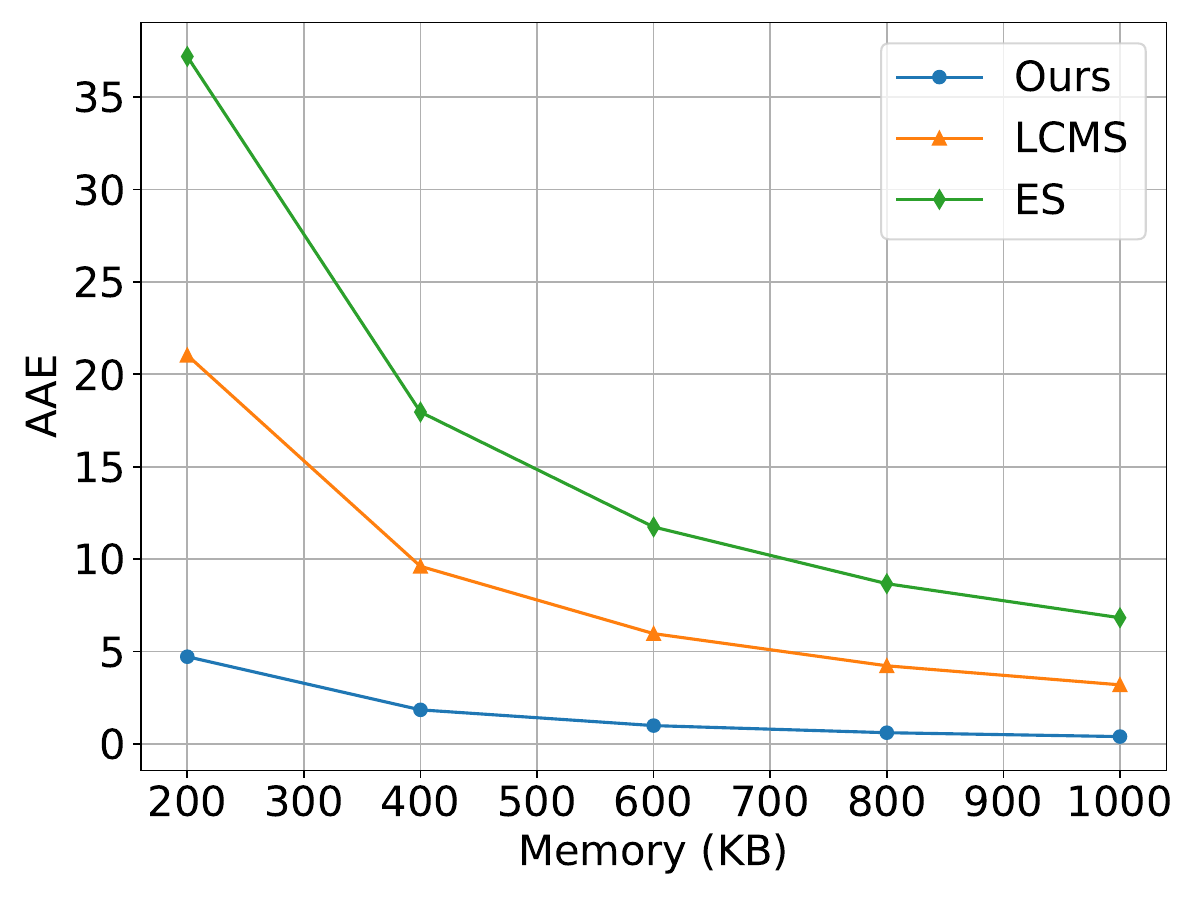}
    \label{eva:fs_mawi:aae}
    }
    \subfigure[ARE on IMC DC.]{
    \includegraphics[width=0.225\textwidth]{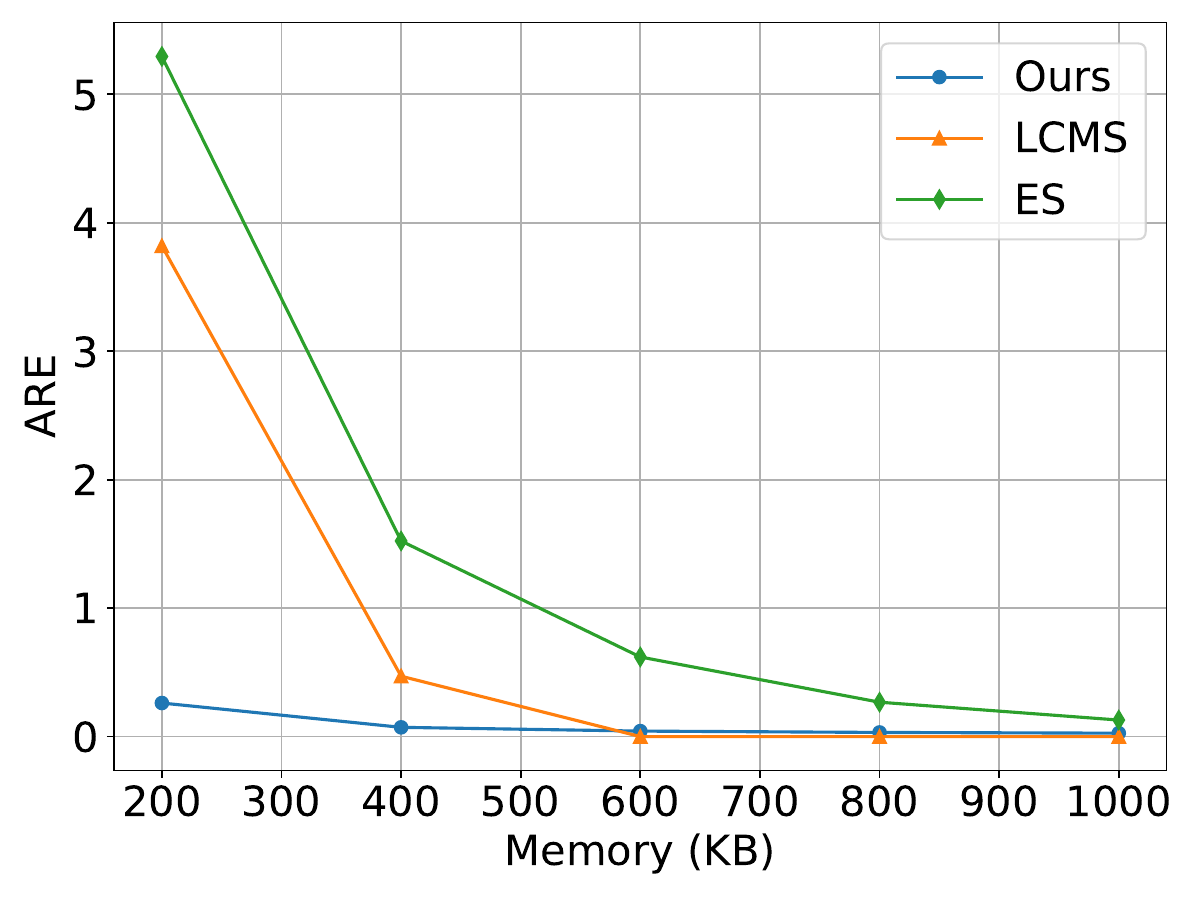}
    \label{eva:fs_imcdc:are}
    }
    \subfigure[AAE on IMC DC.]{
    \includegraphics[width=0.225\textwidth]{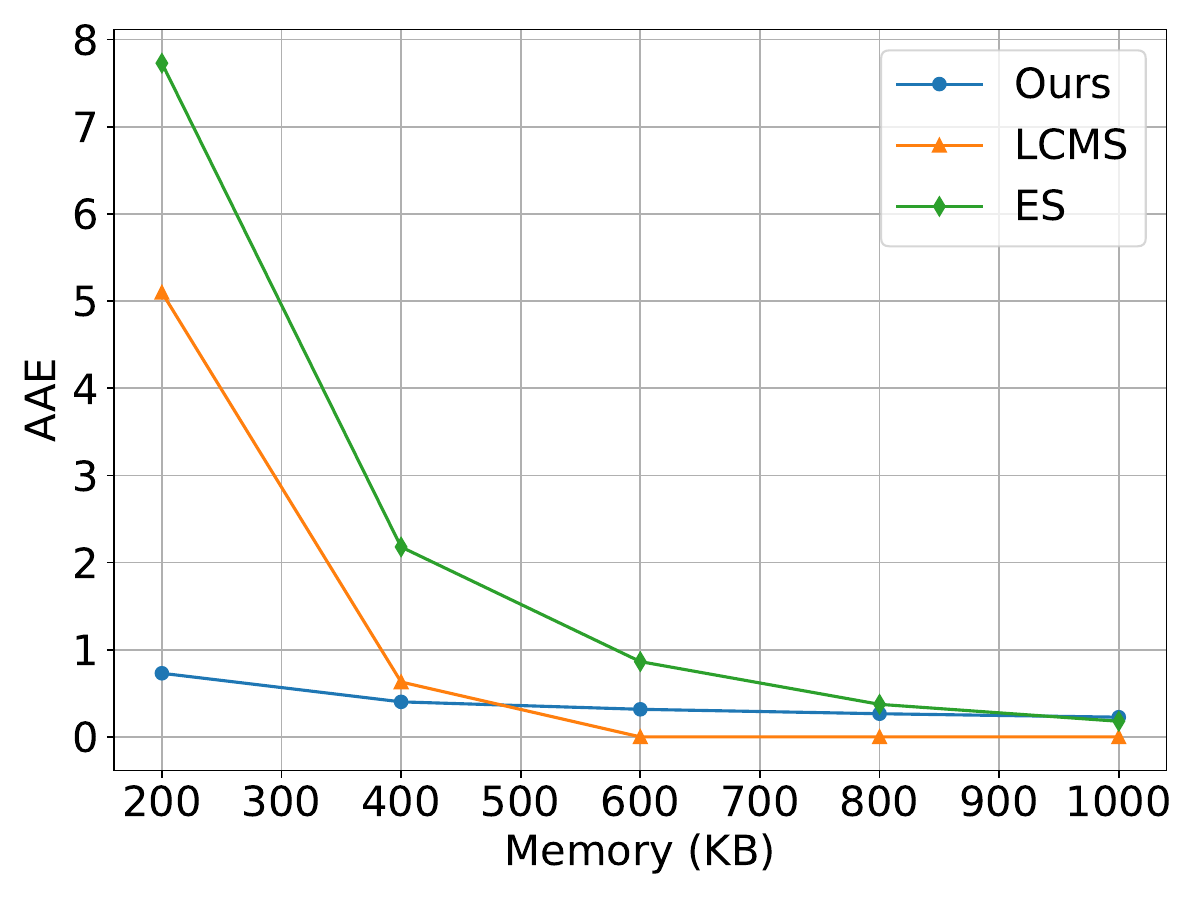}
    \label{eva:fs_imcdc:aae}
    }
    \caption{Accuracy of flow size query on other datasets. \textnormal{\textit{(When memory is 200KB, the ARE and AAE of MS on MAWI are 224 and 271, respectively, those on IMC DC are 29 and 108, respectively.)}}}
    \label{eva:fs_other}
\end{figure*}

\begin{figure*}[t!]
    \centering
    \subfigure[F1 on MAWI.]{
    \includegraphics[width=0.225\textwidth]{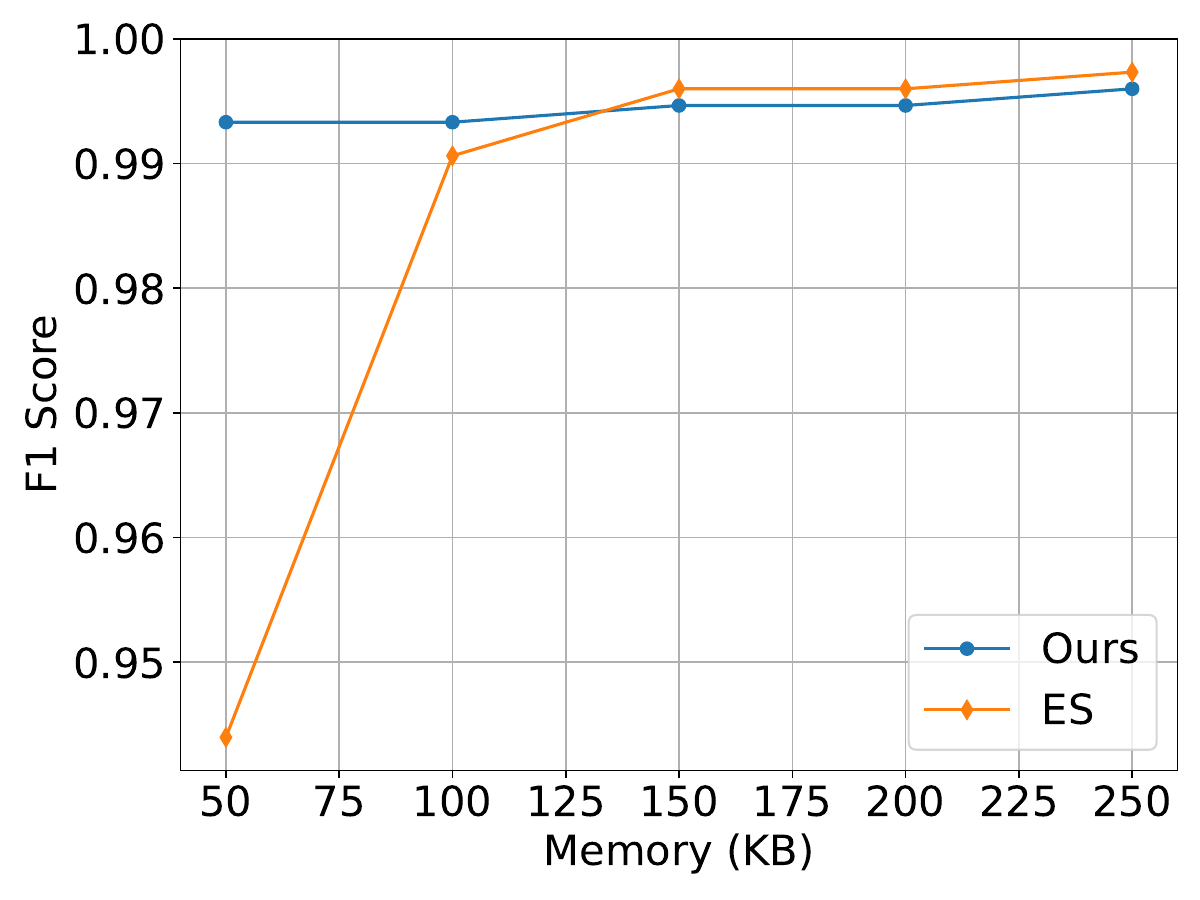}
    \label{eva:hh_mawi:f1}
    }
    \subfigure[ARE on MAWI.]{
    \includegraphics[width=0.225\textwidth]{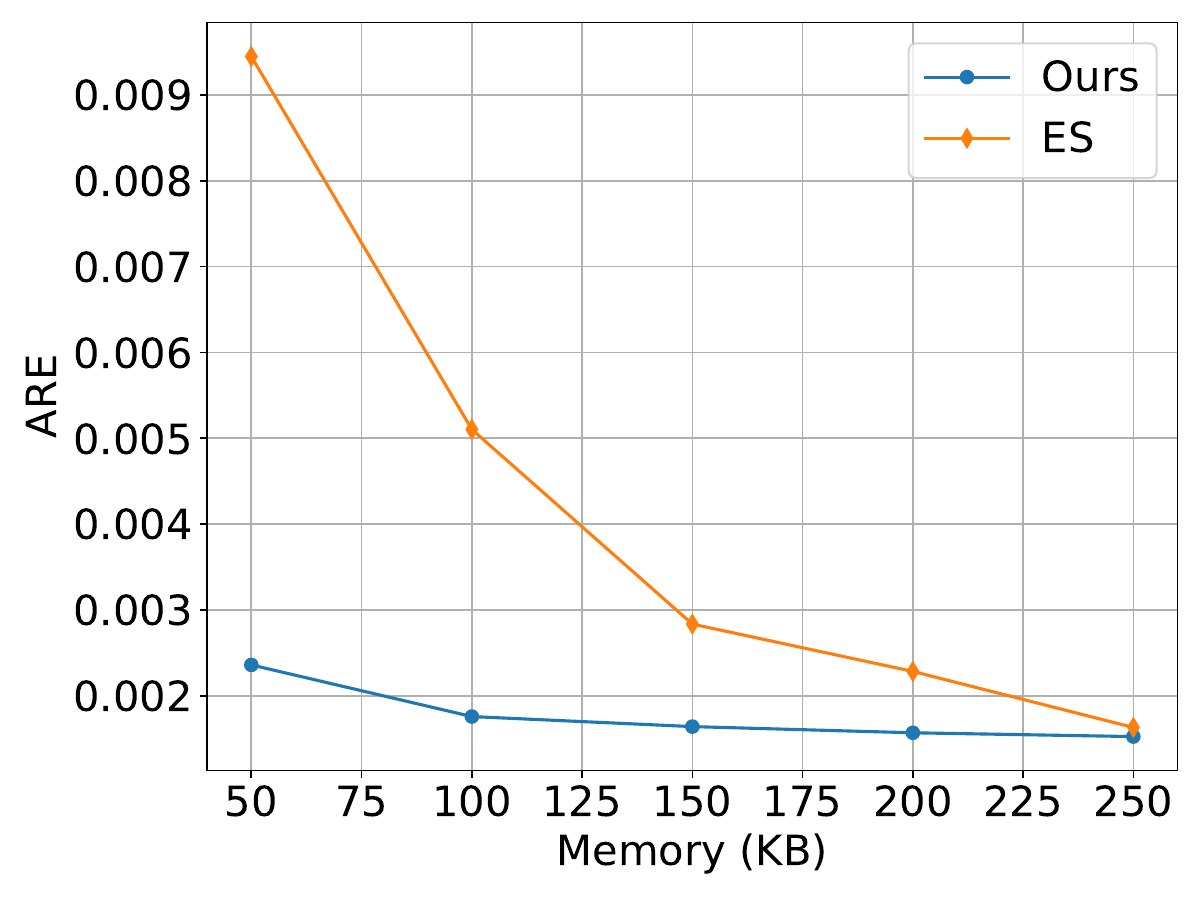}
    \label{eva:hh_mawi:are}
    }
    \subfigure[F1 on IMC DC.]{
    \includegraphics[width=0.225\textwidth]{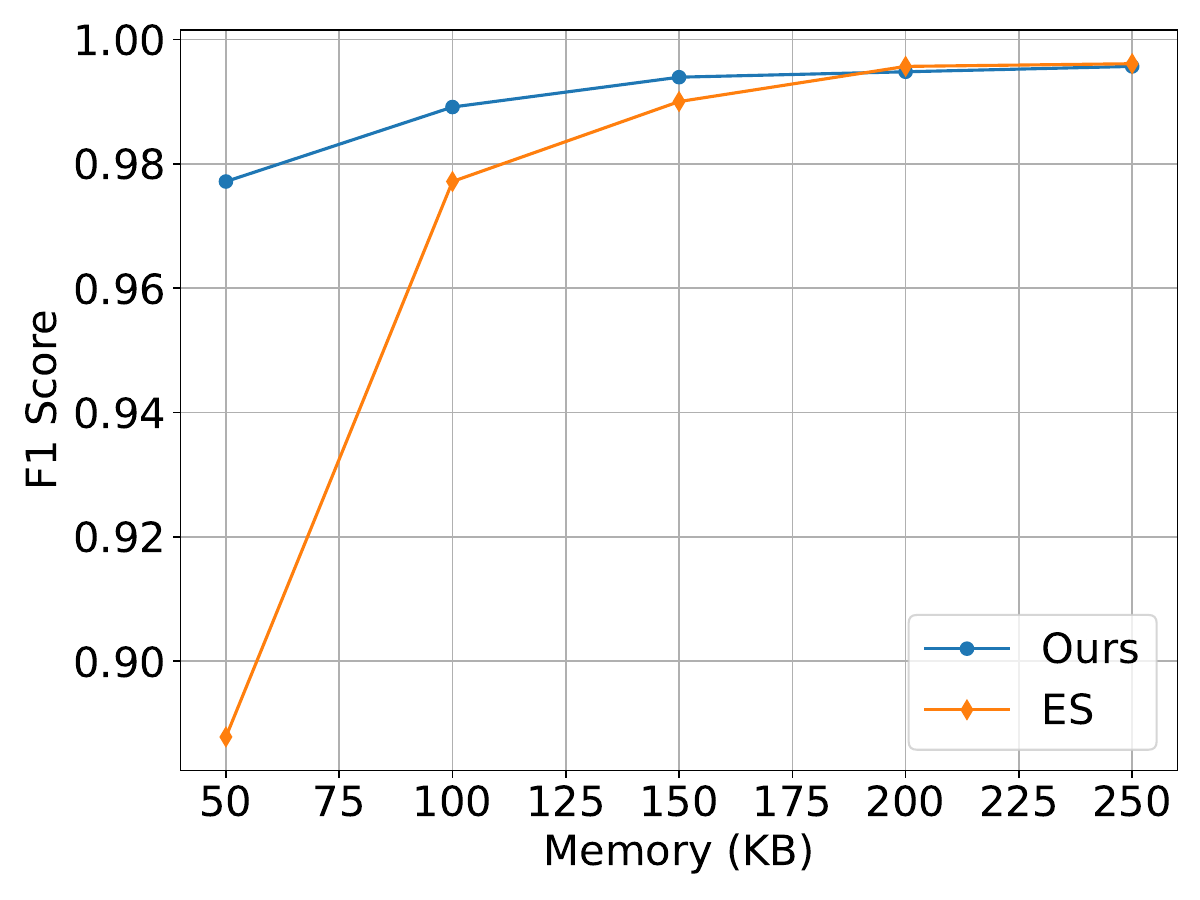}
    \label{eva:hh_imcdc:f1}
    }
    \subfigure[ARE on IMC DC.]{
    \includegraphics[width=0.225\textwidth]{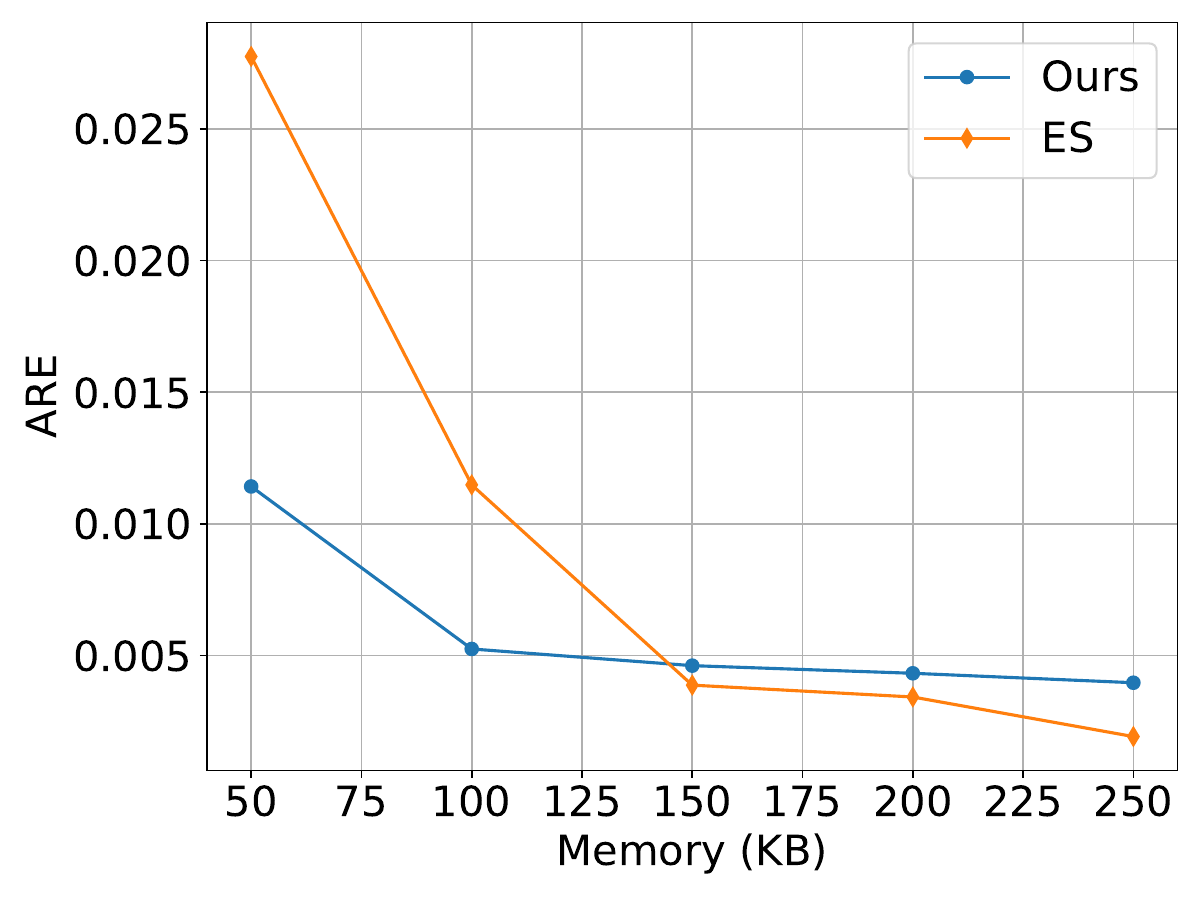}
    \label{eva:hh_imcdc:are}
    }
    \caption{Accuracy of flow size query on other datasets.}
    \label{eva:hh_other}
\end{figure*}

\bbb{Accuracy of flow size query (Figure~\ref{eva:fs}):}
We compare \alg{} with LCMS, MS, and ES. \textit{We find that \alg{} achieves the highest accuracy among all 4 algorithms.} On average, the ARE of \alg{} is 11.8 and 18.8 times lower than that of LCMS and ES, respectively, while its AAE is also 8.1 and 12.1 lower than those of LCMS and ES, respectively.
It is worth noting that meta-sketch is not practical at our current data scale for two main reasons:
\begin{enumerate}[leftmargin=*]
    \item When the memory budget is set to 200 KB, training a 500k-step model fails to converge, leading to poor accuracy.

    \item When the memory is increased to 400 KB, meta-sketch requires more than 24 GB of GPU memory for training.
\end{enumerate}

\bbb{Accuracy of heavy hitter query (Figure~\ref{eva:hh}):}
We compare \alg{} with ES. \textit{We find that \alg{} achieves higher accuracy among the 2 algorithms.} Under a memory budget of 50KB, \alg{} reaches an F1 score of 0.94, whereas the F1 score of ES is 0.74. In addition, the ARE of \alg{} is also on average 2.6 times lower than that of ES.

\bbb{Accuracy of HHH query (Figure~\ref{eva:hhh}):}
We compare \alg{} with CocoS. \textit{We find that \alg{} achieves higher accuracy among the 2 algorithms.} Under a memory budget of 50KB, \alg{} reaches an F1 score of 0.94, whereas the F1 score of CocoS is 0.82. In addition, the ARE of \alg{} is also on average 1.9 times lower than that of ES.

\bbb{Accuracy on other datasets (Figure~\ref{eva:fs_other}-\ref{eva:hh_other}:)}
Apart from the CAIDA dataset, we evaluate the accuracy on flow size query and heavy hitter query on two additional datasets and find that \alg{} achieves high accuracy.
For flow size query, on average, \alg{}’s ARE is 5.3 and 13.2 times lower than those of LCMS and ES, respectively, while its AAE is also 3.9 and 8.2 lower than those of LCMS and ES, respectively.
For heavy hitter query, under a memory budget of 50KB, \alg{} reaches an F1 score of 0.99 and 0.98 on the two datasets, respectively, whereas the F1 score of ES is 0.94 and 0.89, respectively.
It is worth noting that, on the IMC DC dataset, when the memory size is large, ElasticSketch achieves higher accuracy. This is because, with ample memory, ElasticSketch’s heavy part has sufficient cells to store and identify potential large flows, thus avoiding errors from misclassifying large flows. In this scenario, \alg{} also maintains high accuracy (F1 score > 0.99, ARE < 0.05).

\subsection{Micro Benchmark}

\begin{figure}[!t]
    \centering  
    \includegraphics[width=0.25\textwidth]{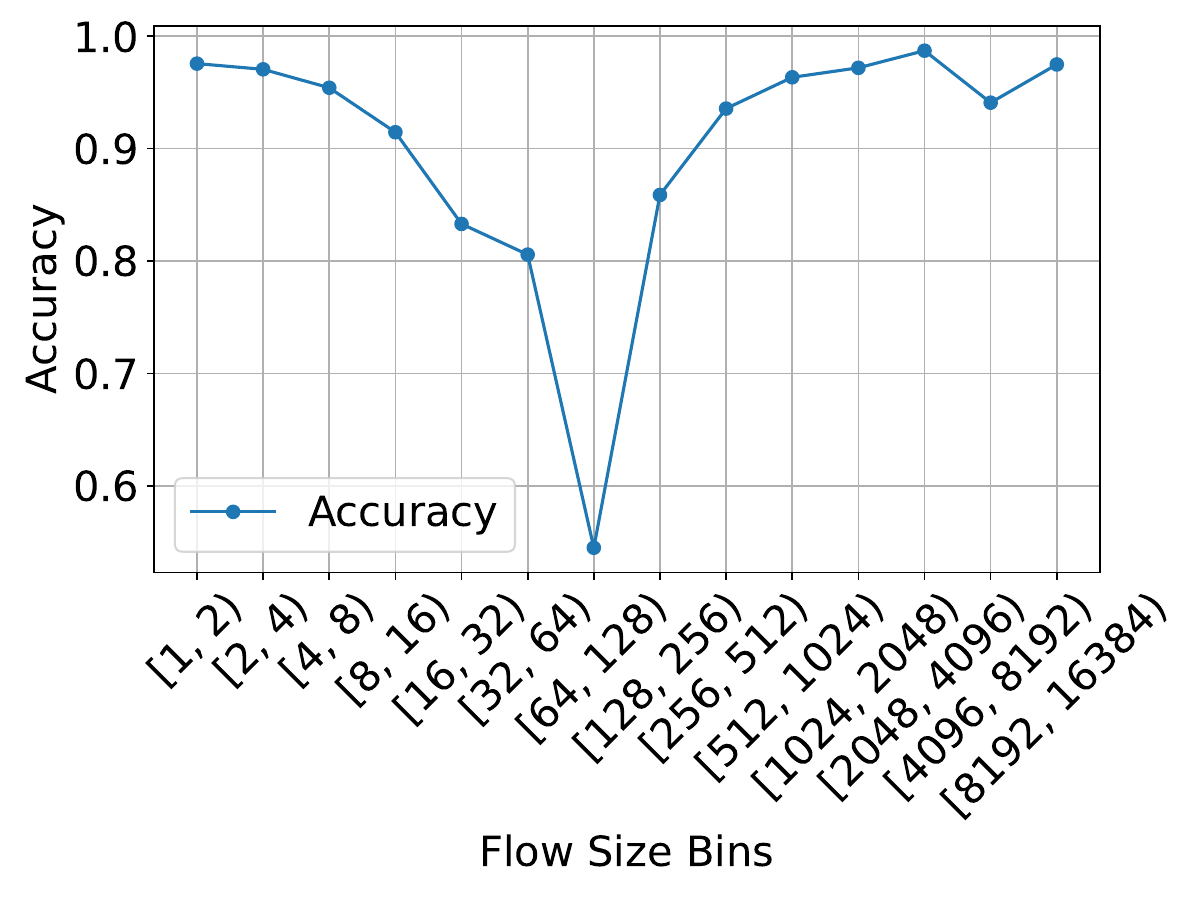}
    \caption{Model accuracy.}
    \label{eva:model}
\end{figure}
\begin{figure}[t!]
    \centering
    \subfigure[F1 score of model.]{
    \includegraphics[width=0.225\textwidth]{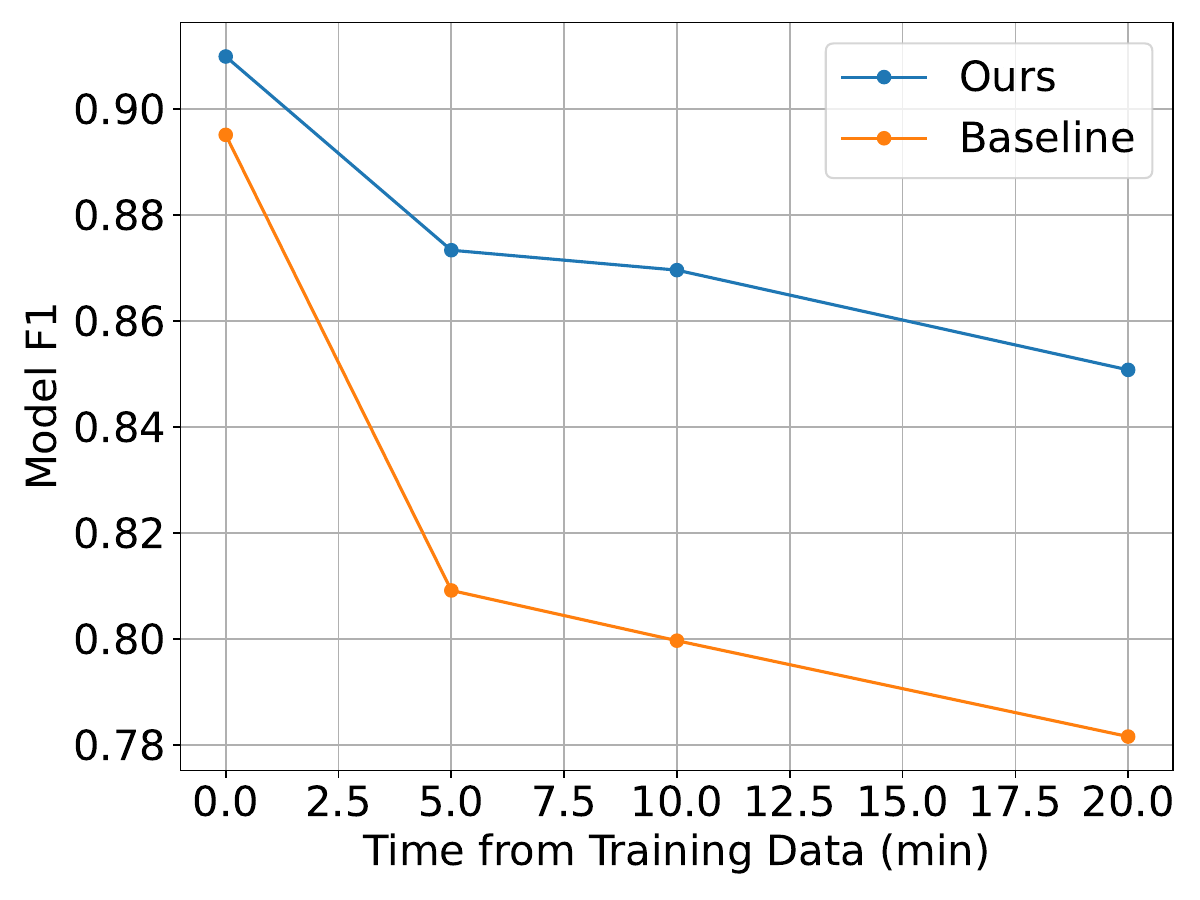}
    \label{eva:time:model}
    }
    \subfigure[F1 score of sketch.]{
    \includegraphics[width=0.225\textwidth]{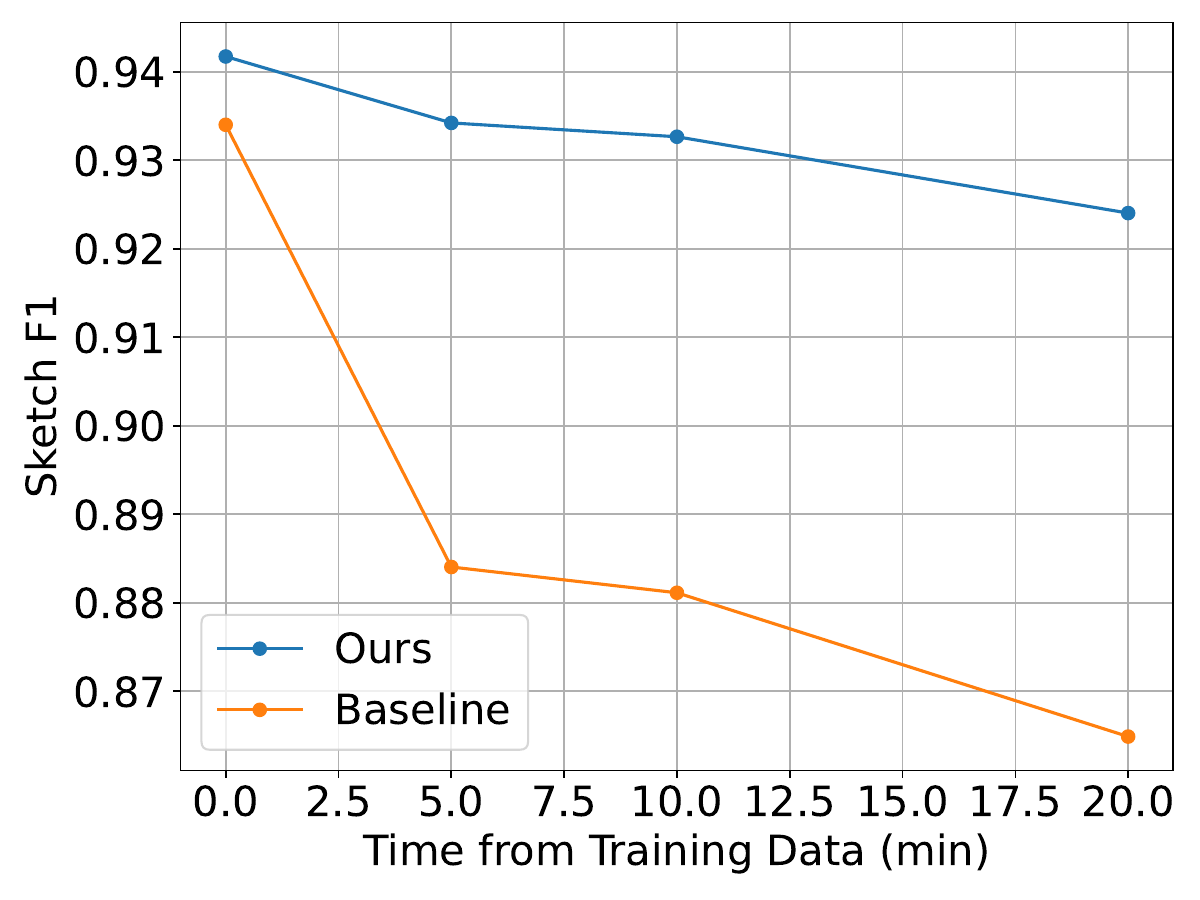}
    \label{eva:time:sketch}
    }
    \caption{Accuracy over time.}
    \label{eva:time}
\end{figure}

\bbb{Model accuracy (Figure~\ref{eva:model}):} We evaluate the classification accuracy for flows of different sizes and observe that \alg{} accurately classifies both very large and very small flows, which meets our expectations. For flows smaller than 16 packets and those larger than 256 packets, the model’s classification accuracy exceeds 90\%. Although accuracy for flows within the [32,64) range is relatively low, it has little impact on the overall performance and is thus acceptable.

\bbb{Enhancement by Using Full Packet Headers (Figure~\ref{eva:time}).}
We compare the accuracy of \alg{} with a baseline algorithm that uses only flow IDs for classification. We train the classifier on a training set and test it on 4 test sets, each collected at a different time interval from the training set. We also evaluate the accuracy of heavy hitter query under 50KB. \textit{We find that when using full packet headers, the classifier can accurately infer flow sizes over an extended period, thereby preserving the sketch’s high accuracy.}
As shown in Figure~\ref{eva:time}, although at minute 0 the baseline and \alg{} yield essentially the same results, the baseline’s classifier accuracy and its sketch accuracy decline rapidly over time, whereas \alg{}’s accuracy shows only a slight decrease. At the 20-minute mark, the classifier’s F1 scores for \alg{} and the baseline drop by 0.059 and 0.113, respectively, while the sketch’s end-to-end F1 scores also drop by 0.017 and 0.069, respectively.

\section{Related Work}

\bbb{Traditional sketches:}
Sketches can broadly be categorized into two types:
\textit{1) Classic sketches} consist of a counter matrix and multiple hash functions. During updates and queries, flow IDs are hashed into multiple counters, and the mapped counters are then updated and queried accordingly. Typical classic sketches include the Count-Min Sketch (CMS) \cite{cmsketch}, the Conservative Update Sketch (CUS) \cite{cusketch}, and the Count Sketch (CS) \cite{csketch}. However, classic sketches fail to account for the highly skewed nature of network traffic, resulting in memory waste.
\textit{2) Sophisticated sketches} address this problem by separating large flows from small flows \cite{elasticsketch, sketchlearn, nitrosketch, nze-sketch, bitsense}. These sketches typically consist of multiple parts, with different parts using different data structures to record flows of varying sizes. A typical sophisticated sketch is ElasticSketch \cite{elasticsketch}, which is composed of a heavy part and a light part. The heavy part is a key-value table, while the light part is a CM sketch. Packets are first inserted into the heavy part. When the heavy part becomes full, ElasticSketch uses an eviction method to remove the flow that is most likely to be small and inserts it into the light part. Researchers have attempted to improve sketch accuracy by adjusting the flow selection method for eviction, but most of these attempts have been based on experience.
Additional sketches have been proposed for specialized tasks, such as heavy hitter query \cite{css, mvsketch, precision}, hierarchical heavy hitter query \cite{rhhh, cocosketch}, and DDoS victim/super-spreader query \cite{opensketch, spreadsketch}.

\bbb{ML-based sketches:}
In recent years, researchers have attempted to use machine learning methods to improve sketch performance. Learned Count-Min Sketch (LCMS) \cite{lcmsketch} employs an RNN to learn and infer whether a flow is large and uses an additional hash table to record large flows. Other solutions \cite{bertsimas2021frequency, mlsketch, metasketch} use ML to enhance hashing, updating, and querying processes within the sketch, leading to improved performance. The main difference between these works and ours is that they only learn features from the flow ID and distribution, without utilizing other information carried by the packets.

\bbb{LLM for other network tasks:}
Some works explore how to adapt LLMs to other network operations, such as traffic classification, viewport prediction, adaptive bitrate streaming, and cluster job scheduling. Typical works include PERT \cite{pert}, ET-BERT \cite{et-bert}, YaTC \cite{yatc}, and NetLLM \cite{netllm}.

\section{Conclusion}

In this paper, we propose \alg{}, which combines a novel two-part data structure with an LLM-powered flow classifier to effectively separate large flows from small ones and reduce collisions. By employing a soft-label mechanism for real-time flow predictions, it accurately identifies potential heavy flows while minimizing misclassifications. Experimental results on real-world datasets confirm that \alg{} achieves a significant \(7.5\times\) improvement in accuracy over state-of-the-art methods, demonstrating its versatility for diverse network stream mining tasks.
All related code is open-sourced on Github.

% email: llm-sketch@outlook.com
% github: LLM-Sketch
% password: lab1806@PKU

\bibliographystyle{ACM-Reference-Format}
\bibliography{reference}

\end{document}